\documentclass[aps, prb, twocolumn, reprint, superscriptaddress, floatfix, citeautoscript, longbibliography]{revtex4-1}
\usepackage[T1]{fontenc} 
\usepackage[utf8]{inputenc}
\usepackage{newtxtext}
\usepackage{newtxmath}
\usepackage{graphicx}
\usepackage{siunitx}
\usepackage[version=4]{mhchem}
\usepackage{booktabs}
\usepackage[english]{babel}
\usepackage{csquotes}
\usepackage{microtype}
\usepackage[unicode,colorlinks=true,allcolors=blue]{hyperref}
\usepackage{cleveref}

\newcommand{\nmr}{NMR}

\newcommand{\bnmr}{$\beta$-NMR}

\newcommand{\slr}{SLR}

\newcommand{\rf}{RF}

\newcommand{\cw}{CW}

\newcommand{\eli}{\ce{^{8}Li}}
\newcommand{\elip}{\ce{^{8}Li^{+}}}

\newcommand{\lle}{$\lambda_e$}
\newcommand{\ldiff}{$\lambda_{\mathrm{diff}}$}

\newcommand{\latin}[1]{\emph{#1}}

\begin{document} 

\title{Ionic and electronic properties of the topological insulator \ce{Bi2Te2Se} investigated using $\beta$-detected nuclear magnetic relaxation and resonance of \ce{^{8}Li}}

\author{Ryan M. L. McFadden}
\email[Email: ]{rmlm@chem.ubc.ca}
\affiliation{Department of Chemistry, University of British Columbia, Vancouver, BC V6T~1Z4, Canada}
\affiliation{Stewart Blusson Quantum Matter Institute, University of British Columbia, Vancouver, BC V6T~1Z4, Canada}

\author{Aris Chatzichristos}
\affiliation{Stewart Blusson Quantum Matter Institute, University of British Columbia, Vancouver, BC V6T~1Z4, Canada}
\affiliation{Department of Physics and Astronomy, University of British Columbia, Vancouver, BC V6T~1Z1, Canada}

\author{Kim H. Chow}
\affiliation{Department of Physics, University of Alberta, 4-181 CCIS, Edmonton, AB T6G~2E1, Canada}

\author{David L. Cortie}
\altaffiliation{Current address: Institute for Superconducting and Electronic Materials, Australian Institute for Innovative Materials, University of Wollongong, North Wollongong, NSW 2500, Australia}
\affiliation{Department of Chemistry, University of British Columbia, Vancouver, BC V6T~1Z4, Canada}
\affiliation{Stewart Blusson Quantum Matter Institute, University of British Columbia, Vancouver, BC V6T~1Z4, Canada}
\affiliation{Department of Physics and Astronomy, University of British Columbia, Vancouver, BC V6T~1Z1, Canada}

\author{Martin H. Dehn}
\affiliation{Stewart Blusson Quantum Matter Institute, University of British Columbia, Vancouver, BC V6T~1Z4, Canada}
\affiliation{Department of Physics and Astronomy, University of British Columbia, Vancouver, BC V6T~1Z1, Canada}

\author{Derek Fujimoto}
\affiliation{Stewart Blusson Quantum Matter Institute, University of British Columbia, Vancouver, BC V6T~1Z4, Canada}
\affiliation{Department of Physics and Astronomy, University of British Columbia, Vancouver, BC V6T~1Z1, Canada}

\author{Masrur D. Hossain}
\altaffiliation{Current address: ASML US, Inc, 399 W Trimble Rd, San Jose, CA 95131, USA}
\affiliation{Department of Physics and Astronomy, University of British Columbia, Vancouver, BC V6T~1Z1, Canada}

\author{Huiwen Ji}
\altaffiliation{Current address: Department of Materials Science and Engineering, University of California, Berkeley, Berkeley, CA 94720, USA}
\affiliation{Department of Chemistry, Princeton University, Princeton, New Jersey 08544, USA}

\author{Victoria L. Karner}
\affiliation{Department of Chemistry, University of British Columbia, Vancouver, BC V6T~1Z4, Canada}
\affiliation{Stewart Blusson Quantum Matter Institute, University of British Columbia, Vancouver, BC V6T~1Z4, Canada}

\author{Robert F. Kiefl}
\affiliation{Stewart Blusson Quantum Matter Institute, University of British Columbia, Vancouver, BC V6T~1Z4, Canada}
\affiliation{Department of Physics and Astronomy, University of British Columbia, Vancouver, BC V6T~1Z1, Canada}
\affiliation{TRIUMF, 4004 Wesbrook Mall, Vancouver, BC V6T~2A3, Canada}

\author{C. D. Philip Levy}
\affiliation{TRIUMF, 4004 Wesbrook Mall, Vancouver, BC V6T~2A3, Canada}

\author{Ruohong Li}
\affiliation{TRIUMF, 4004 Wesbrook Mall, Vancouver, BC V6T~2A3, Canada}

\author{Iain McKenzie}
\affiliation{TRIUMF, 4004 Wesbrook Mall, Vancouver, BC V6T~2A3, Canada}
\affiliation{Department of Chemistry, Simon Fraser University, Burnaby, BC V5A~1S6, Canada}

\author{Gerald D. Morris}
\affiliation{TRIUMF, 4004 Wesbrook Mall, Vancouver, BC V6T~2A3, Canada}

\author{Oren Ofer}
\affiliation{TRIUMF, 4004 Wesbrook Mall, Vancouver, BC V6T~2A3, Canada}

\author{Matthew R. Pearson}
\affiliation{TRIUMF, 4004 Wesbrook Mall, Vancouver, BC V6T~2A3, Canada}

\author{Monika Stachura}
\affiliation{TRIUMF, 4004 Wesbrook Mall, Vancouver, BC V6T~2A3, Canada}

\author{Robert J. Cava}
\affiliation{Department of Chemistry, Princeton University, Princeton, New Jersey 08544, USA}

\author{W. Andrew MacFarlane}
\email[Email: ]{wam@chem.ubc.ca}
\affiliation{Department of Chemistry, University of British Columbia, Vancouver, BC V6T~1Z4, Canada}
\affiliation{Stewart Blusson Quantum Matter Institute, University of British Columbia, Vancouver, BC V6T~1Z4, Canada}
\affiliation{TRIUMF, 4004 Wesbrook Mall, Vancouver, BC V6T~2A3, Canada}

\date{\today}

\begin{abstract}
We report measurements on the high temperature ionic and low temperature electronic properties of the 3D topological insulator \ce{Bi2Te2Se} using ion-implanted \ce{^{8}Li} $\beta$-detected nuclear magnetic relaxation and resonance. With implantation energies in the range \SIrange[range-phrase=--,range-units=single]{5}{28}{\kilo\electronvolt}, the probes penetrate beyond the expected range of the topological surface state, but are still within \SI{250}{\nano\meter} of the surface.
At temperatures above \SI{\sim 150}{\kelvin}, spin-lattice relaxation measurements reveal isolated \ce{^{8}Li^{+}} diffusion with an activation energy $E_{A} = \SI{0.185 \pm 0.008}{\electronvolt}$ and attempt frequency $\tau_{0}^{-1} = \SI{8 \pm 3 e11}{\per\second}$ for atomic site-to-site hopping.
At lower temperature, we find a linear Korringa-like relaxation mechanism with a field dependent slope and intercept, which is accompanied by an anomalous field dependence to the resonance shift.
We suggest that these may be related to a strong contribution from orbital currents or the magnetic freezeout of charge carriers in this heavily compensated semiconductor, but that conventional theories are unable to account for the extent of the field dependence.
Conventional NMR of the stable host nuclei may help elucidate their origin.
\end{abstract}

\maketitle

\section{Introduction \label{sec:introduction}}

\begin{figure}
\centering
\includegraphics[trim={0.9cm 1.1cm 0 0},clip,width=1.0\columnwidth]{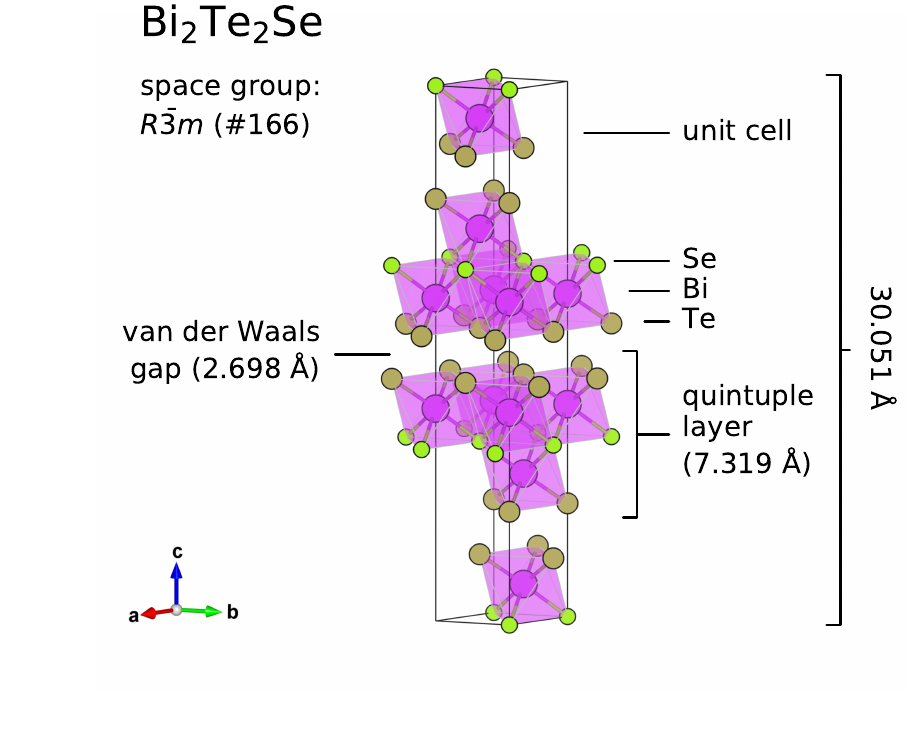}
\caption{ \label{fig:bts-unit-cell}
Crystal structure of \ce{Bi2Te2Se}~\cite{2011-Jia-PRB-84-235206}, consisting of \ce{Te}-\ce{Bi}-\ce{Se}-\ce{Bi}-\ce{Te} layers.
These quintuple layers are weakly coupled through van der Waals interactions, giving rise to a characteristic gap between adjacent \ce{Te} planes.
This atomic arrangement is analogous to transition metal dichalcogenides, where it is possible to insert foreign atoms and small molecules within
the van der Waals gap.
The structure was drawn using \texttt{VESTA}~\cite{vesta}.
}
\end{figure}

Bismuth chalcogenides with the formula \ce{Bi2Ch3} ($\ce{Ch}=\ce{S,Se,Te}$) are narrow gap semiconductors that have been studied for decades for their thermoelectric properties.
They crystallize in the layered tetradymite structure~\cite{2013-Cava-JMCC-1-3176, 2017-Heremans-NRM-2-17049}, consisting of stacks of strongly bound \ce{Ch}-\ce{Bi}-\ce{Ch}-\ce{Bi}-\ce{Ch} quintuple layers (QLs) loosely coupled by van der Waals (vdW) interactions (see \Cref{fig:bts-unit-cell}).
More recently, interest in their electronic properties has exploded~\cite{2010-Hasan-RMP-82-3045} following the realization that strong spin-orbit coupling and band inversion combine to make them 3D topological insulators (TIs)~\cite{2011-Hasan-ARCMP-2-55}, characterized by a gapless topological surface state (TSS).
Electronically, this family of TIs is characterized by a relatively insulating bulk and a robustly conductive surface, with greater contrast in conductivity between the two regions significantly facilitating identification and study.
The prevalence for self-doping in binary chalcogenides (\latin{e.g.}, \ce{Bi2Se3} or \ce{Bi2Te3}) often yields crystals far from insulating in the bulk, masking the signature of the conductive surface state.
This has been mitigated, for example, in the most widely studied tetradymite TI \ce{Bi2Se3} with \ce{Ca} doping to suppress the more usual $n$-type conductivity~\cite{2009-Hor-PRB-79-195208, 2009-Hsieh-N-460-1101}.
On the other hand, the stoichiometric ordered~\cite{1963-Nakajima-JPCS-24-479, 2011-Jia-PRB-84-235206} ternary line compound \ce{Bi2Te2Se} (BTS) exhibits a much lower conductivity thanks to its fortuitous crystal chemistry~\cite{2011-Jia-PRB-84-235206}, with anti-site defects (\latin{e.g.}, bismuth substitution on a tellurium site) playing an important role~\cite{2012-Scanlon-AM-24-2154}.
Indeed, BTS crystals with a characteristic bulk band gap of \SI{\sim 0.3}{\electronvolt}~\cite{2012-Akrap-PRB-86-235207} have demonstrated such desired large bulk resistivities at low temperatures~\cite{2010-Ren-PRB-82-241306, 2011-Jia-PRB-84-235206, 2012-Xiong-PE-44-917, 2012-Xiong-PRB-86-045314}.
A great deal is known about its surface properties, with angle-resolved photoemission spectroscopy (ARPES) revealing the characteristic linear dispersion about the Dirac point~\cite{2012-Neupane-PRB-85-235406, 2012-Scanlon-AM-24-2154, 2012-Miyamoto-PRL-109-166802, 2015-Nurmamat-PRB-88-081301, 2015-Neupane-PRL-115-116801, 2017-Frantzeskakis-PRX-7-041041, 2018-Munisa-PRB-97-115303};
however, the material is not an ideal TI due to the close proximity of the Dirac point to the top of the bulk valence band~\cite{2017-Heremans-NRM-2-17049}.

With the novel electronic structure of the \ce{Bi} chalcogenides evident primarily in the TSS, much inquiry has focused on surface sensitive probes, such as ARPES, scanning tunnelling spectroscopy (STS), and transport.
On the other hand, NMR is well-known to reveal electronic properties in metals through hyperfine coupling of the nucleus to surrounding electron spins that gives rise to the Knight shift, a measure of the Pauli spin susceptibility, and the Korringa spin-lattice relaxation (SLR)~\cite{1950-Korringa-P-16-601, 1990-Slichter-PMR}.
Theory predicts dramatic effects in such quantities for the TSS~\cite{2012-Vazifeh-PRB-86-045451}, but NMR is generally a bulk probe with very little sensitivity to the surface.
Despite this, a considerable body of conventional NMR in the \ce{Bi2Ch3} TIs has accumulated.
Since these results are closely related to the present study, we give a brief summary.

All elements in BTS have NMR-active isotopes.
The most conspicuous feature of \SI{100}{\percent} abundant \ce{^{209}Bi} NMR is the strong quadrupolar interaction.
However, the broad quadrupole pattern shows clear evidence of a shift related to carrier density, indicating a very strong hyperfine coupling~\cite{2013-Nisson-PRB-87-195202}, confirmed more recently by very high field NMR~\cite{2015-Mukhopadhay-PRB-91-081105}.
The behavior of $1/T_1$ is less consistent, with results ranging from $T$-linear to nearly $T$-independent~\cite{2012-Young-PRB-86-075137, 2013-Nisson-PRB-87-195202} at low temperature.
The low abundance (\SI{\sim 7}{\percent}) pure magnetic spin $1/2$ probes \ce{^{77}Se} and \ce{^{125}Te} show a small $T$-independent shift, and Korringa relaxation at low temperature in orientationally averaged powder spectra~\cite{2014-Koumoulis-AFM-24-1519}.
This Korringa relaxation is enhanced in nanocrystals and was attributed to the TSS~\cite{2013-Koumoulis-PRL-110-026602}, but 
given the evidence for aging effects that stabilize a conventional metallic surface state~\cite{2010-Bianchi-NC-1-128},
this connection remains unclear. More recent work in single crystals has identified distinct resonances from inequivalent chalcogen planes, and a more detailed analysis of the chalcogen NMR is required~\cite{2015-Podorozhkin-PSS-57-1741,2016-Georgieva-PRB-93-195120,2017-Antonenko-PSS-59-2331}.

While a great deal is known about the TSS from surface sensitive probes, little is known about how this behaviour transitions to the bulk as a function of depth below the crystal surface.
We plan to address this question using highly spin-polarized radioactive ions, specifically, using depth-resolved $\beta$-detected nuclear magnetic resonance and relaxation ($\beta$-NMR)~\cite{2015-MacFarlane-SSNMR-68-1} with short-lived \ce{^{8}Li} as the probe nucleus, similar to low-energy muon spin rotation ($\mu$SR)~\cite{2004-Bakule-CP-45-203}.
As a first step, here we present results using relatively high implantation energies, \SIrange[range-phrase=--,range-units=single]{5}{28}{\kilo\electronvolt}, corresponding to \ce{^{8}Li^{+}} implantation depths below the TSS.
We find that below \SI{\sim 150}{\kelvin} \ce{^{8}Li} experiences strongly field dependent relaxation and resonance shifts.
The relaxation is $T$-linear and reminiscent of a Korringa mechanism, but conventional theory for NMR cannot account for its field dependence, which we suggest is due to magnetic carrier freeze out.
At higher temperatures, an additional relaxation mechanism is observed, originating from ionic diffusion of isolated \ce{^{8}Li^{+}} in the vdW gap.

The remainder of the paper is organized as follows:
a brief description of the experiment is given in \Cref{sec:experiment},
followed by the results and analysis in \Cref{sec:results}.
A detailed discussion is given in \Cref{sec:discussion}, focusing on
the high temperature dynamics of the isolated \ce{^{8}Li^{+}} ion in the vdW gap (\Cref{sec:discussion:diffusion}), and the electronic properties of BTS giving rise to the field dependent relaxation and resonance shifts at low temperature (\Cref{sec:discussion:electronic}).
Finally, a concluding summary can be found in \Cref{sec:conclusion}.

\section{Experiment \label{sec:experiment}}

Since implanted ion \bnmr\ is not a widely known technique, here we summarize some of its main features.
More detailed accounts can be found in Refs.~\onlinecite{1983-Ackermann-TCP-31-291, 2015-MacFarlane-SSNMR-68-1}.

\subsection{The \bnmr\ Technique \label{sec:experiment:technique}}

In many ways, \bnmr\ is very similar to stable isotope NMR~\cite{1990-Slichter-PMR}.
The nuclear spin senses the local magnetic fields via the Zeeman interaction and their time-averages contribute to the resonance shift and lineshape.
In addition, since \eli\ possesses a non-zero nuclear quadrupole moment, the nuclear spin is coupled to the local electric field gradient (EFG)~\cite{1990-Slichter-PMR, 1957-Cohen-SSP-5-321},
\begin{equation*}
   eq = \frac{\partial^2 V}{\partial x_i \partial x_j},
\end{equation*}
a tensor that is zero under cubic symmetry.
When the EFG tensor is non-vanishing, the quadrupolar interaction splits the resonance into a set of $2I$ satellites.
The integer spin $(I=2)$ of \eli\ (uncommon in conventional NMR) has the important consequence that the quadrupolar spectrum has no ``main line'' at the Larmor frequency,
\begin{equation*} \label{eq:larmor}
   \omega_{0} = 2 \pi \nu_{0} = \gamma B_{0},
\end{equation*}
determined by the gyromagnetic ratio $\gamma$ of the NMR nucleus and the (dominant) applied magnetic field $B_{0}$.
In contrast, for the more familiar case of half-integer $I$, the $m = \pm 1/2$ transition yields, to first order, a line at $\nu_{0}$ unperturbed by quadrupole effects~\cite{1957-Cohen-SSP-5-321}.
While the quadrupole interaction is often the most important perturbation to the nuclear spin energy levels in nonmagnetic materials, for \eli\ it is still relatively small (in the \si{\kilo\hertz} range), because its nuclear electric quadrupole moment $Q$ is small, compared, for example, to \ce{^{209}Bi}.

Distinct from conventional NMR, the probe is extrinsic to the host, and its lattice site is not known \latin{a priori}.
Like the implanted positive muon $\mu^{+}$ in $\mu$SR, the \elip\ ion generally stops in a high-symmetry site in a crystalline host.
Some site information is available in the resonance spectrum, since the local field and EFG depend on the site, but generally one has to combine this information with knowledge of the structure and calculations to make a precise site assignment.
Moreover, \elip, as a light interstitial, can often be \emph{mobile} near room temperature.
In this case, the local interactions become time dependent, as the probe undergoes stochastic hopping, usually among interstitial sites.
If the average hop rate is near the Larmor frequency, this will cause SLR~\cite{1948-Bloembergen-PR-73-679, 1979-Richards-TCP-15-141, 1988-Beckmann-PR-171-85}.
Since this process is independent of other relaxation mechanisms due to the intrinsic fluctuations of the host, the rates simply add.
At accessible beam intensities the instantaneous number of \eli\ in the sample never exceeds \num{\sim e6}, meaning \eli\ is always present in the ultradilute limit, and interactions between \eli\ can be neglected.
Any diffusive motion is thus characteristic of the \emph{isolated} interstitial.

As with conventional NMR, the SLR is determined by fluctuations at the Larmor frequency in the radio-frequency (\rf) range.
Specific to the \bnmr\ mode of detection, the range of measurable $T_1$ relaxation times is determined by the radioactive lifetime $\tau_\beta$.
As a rule of thumb, measurable $T_1$ values lie in the range $0.01\tau_\beta$ to $100\tau_\beta$~\cite{1983-Ackermann-TCP-31-291}.
Near the upper end of this range, the spin relaxation is very slow and exhibits little or no curvature on the timescale of the measurement.
One can still measure the relaxation rate from the slope, but it is significantly correlated to the initial amplitude of the relaxing polarization signal.

In conventional NMR, the signal-to-noise ratio is proportional to the square of the Larmor frequency, favouring high applied fields.
Consequently, for practical reasons, NMR is often done at a single fixed field in the range of \SI{\sim 10}{\tesla}.
In contrast, the signal in \bnmr\ is independent of frequency, and the field can easily be varied.
This can be useful, for example, in identifying relaxation mechanisms with distinct field dependencies.
In addition, this enables \bnmr\ in the realm of low applied fields (up to \num{10}s of \si{\milli\tesla}).
As the applied field approaches zero, fluctuations of the stable magnetic \emph{nuclei} of the host often become the dominant source of relaxation.
At such low fields, distinction of different nuclei by their Larmor frequency is suppressed, and the isolated \eli\ begins to resonantly lose its spin polarization to the bath of surrounding nuclear spins.
Effectively, this simply appears as another relaxation mechanism active only at low fields.
The extent of the low field regime depends on the moment, density, and NMR properties of the host lattice nuclei.

In many aspects, \bnmr\ is also quite similar to $\mu$SR~\cite{2004-Bakule-CP-45-203}, which too relies on the anisotropic $\beta$-decay from the implanted $\mu^{+}$ to study internal fields in materials.
The two techniques, however, have several important differences, largely due to the properties of their respective probes.
In contrast to \eli, $\mu^{+}$ is:
1) a pure magnetic spin-1/2 probe with no quadrupolar interaction;
2) much shorter lived (lifetime \SI{\sim 2.2}{\micro\second}), making it sensitive to much faster SLR rates;
and 3) chemically unlike an alkali cation, but rather analogous to isolated hydrogen, the basis for much of its use in semiconductors and semimetals~\cite{2009-Cox-RPP-72-116501}.
Consequently, in TIs, $\mu$SR has mainly been used to study magnetically doped~\cite{2017-Krieger-PRB-96-184402, 2018-Duffy-PRB-97-174427} or superconducting~\cite{2018-Leng-PRB-97-054503, 2018-Krieger-JPSCP-21-011028} compositions.
In contrast, the \eli\ relaxation rates we find here are too slow for $\mu$SR, and are in the range of conventional NMR.

\subsection{Measurements \label{sec:experiment:measurement}}

A single crystal of BTS was grown as detailed in Ref.~\onlinecite{2011-Jia-PRB-84-235206} and taken from the more insulating section of a larger boule.
The insulating character was confirmed by resistivity measurements, which show an increase by a factor of \num{\sim 3} below \SI{100}{\kelvin}, with a low temperature maximum \SI{\sim 0.1}{\ohm\centi\meter}, characteristic of the more insulating BTS compositions~\cite{2011-Jia-PRB-84-235206}.
Prior to the \bnmr\ experiments, the crystal, with dimensions \SI[product-units=power]{5 x 4 x 0.1}{\milli\metre}, was cleaved in air and affixed to a sapphire plate using \ce{Ag} paint (SPI Supplies, West Chester, PA) for mounting on a cold finger cryostat.

\bnmr\ experiments were performed at TRIUMF's ISAC facility in Vancouver, Canada.
A low-energy highly polarized beam of \ce{^{8}Li^{+}} was implanted into the BTS single crystal within one of two dedicated spectrometers~\cite{2004-Morris-PRL-93-157601, 2004-Salman-PRB-70-104404, 2015-MacFarlane-SSNMR-68-1, 2014-Morris-HI-225-173}.
The incident \ce{^{8}Li^{+}} ion beam had a typical flux of \SI{\sim e6}{ions\per\second} over a beam spot \SI{\sim 2}{\milli\metre} in diameter.
At the implantation energies $E$ used here (\SIrange{5}{28}{\kilo\electronvolt}), the ions stop at average depths \SI{> 30}{\nano\meter} (see \Cref{fig:srim} in \Cref{sec:implantation}).
The probe nucleus, \ce{^{8}Li}, has nuclear spin $I=2$, gyromagnetic ratio $\gamma/2\pi = \SI{6.3016}{\mega\hertz\per\tesla}$, nuclear electric quadrupole moment $Q=\SI[retain-explicit-plus]{+32.6}{\milli\barn}$, and radioactive lifetime $\tau_{\beta}=\SI{1.21}{\second}$.
Spin-polarization was achieved in-flight by collinear optical pumping with circularly polarized light, yielding a polarization of \SI{\sim 70}{\percent}~\cite{2014-MacFarlane-JPCS-551-012059}, and monitored after ion-implantation through the anisotropic $\beta$-decay emissions of \ce{^{8}Li}.
Specifically, the experimental $\beta$-decay asymmetry (proportional to the average longitudinal spin-polarization) was measured by combining the rates in two opposed scintillation counters~\cite{1983-Ackermann-TCP-31-291, 2015-MacFarlane-SSNMR-68-1}.
The proportionality constant depends on the experimental geometry and the details of the $\beta$-decay, and is \num{\sim 0.1}.

SLR measurements were performed by monitoring the transient decay of spin-polarization both during and following a \SI{4}{\second} pulse of beam~\cite{2006-Salman-PRL-96-147601, 2015-MacFarlane-PRB-92-064409}.
During the pulse, the polarization approaches a steady-state value, while after the pulse, it relaxes to \num{\sim 0}.
Note that the discontinuity at $t=\SI{4}{\second}$ (see \Cref{fig:bts-slr-spectra}) is characteristic of
\bnmr\ \slr\ data.
Unlike conventional \nmr, no \rf\ field is required for the \slr\ measurements, as the probe spins are implanted in a spin state already far from equilibrium.
As a result, it is generally faster and easier to measure \slr\ than the resonance;
however, as a corollary, this type of relaxation measurement has no spectral resolution and represents the spin relaxation of \emph{all} the \eli.
Here, \slr\ rates were measured in small temperature steps from \SIrange[range-phrase=--,range-units=single]{3}{317}{\kelvin} under applied magnetic fields between \SIrange[range-phrase=--,range-units=single]{2.20}{6.55}{\tesla} parallel to the trigonal $c$-axis, and with coarser temperature steps at lower magnetic fields (\SIrange[range-phrase=--,range-units=single]{2.5}{20}{\milli\tesla}) perpendicular to the $c$-axis.
A typical SLR measurement took \SI{\sim 20}{\min}.

Resonances were acquired in a continuous \ce{^{8}Li^{+}} beam with a continuous wave (\cw) transverse \rf\ magnetic field stepped slowly through the \ce{^{8}Li} Larmor frequency.
In this measurement mode, the spin of any on-resonance \ce{^{8}Li} is rapidly precessed by the \rf\ field, resulting in a loss in the average time-integrated asymmetry.
The evolution of the resonance was recorded over a temperature range of \SIrange{3}{317}{\kelvin} in a dedicated high-field spectrometer~\cite{2004-Morris-PRL-93-157601, 2014-Morris-HI-225-173} with a highly-homogeneous magnetic field in the range of \SIrange[range-phrase=--,range-units=single]{2.20}{6.55}{\tesla} parallel to the BTS $c$-axis.
The resonance frequency was calibrated against the position in single crystal \ce{MgO} (100) at \SI{300}{\kelvin}~\cite{2014-MacFarlane-JPCS-551-012033}, with the superconducting solenoid persistent.
Resonance measurements typically took \SI{\sim 30}{\minute} to acquire.

With this mode of measurement, the resonance amplitudes are determined by several factors, some quite distinct from conventional pulsed RF NMR.
First, the maximum amplitude is determined by the baseline asymmetry~\cite{2009-Hossain-PB-404-914}, which represents a time integral of the SLR.
It also depends on the magnitude of the RF magnetic field $B_{1}$ relative to the linewidth, since the RF will only precess \eli\ within a frequency window of width $\sim \gamma B_{1}$.
The resonance amplitude may also be enhanced by slow spectral dynamics occurring up to the second timescale, since the RF is applied at a particular frequency for an integration time of typically \SI{1}{\second}, and any \eli\ that are resonant during this time will be precessed by it.
Quadrupole satellite amplitudes are further reduced by the simple fact that saturating a single quantum transition ($\Delta m = \pm 1$) can, at most, reduce the asymmetry by \SI{25}{\percent}~\cite{2014-MacFarlane-JPCS-551-012059}.
Unsplit resonances can, in contrast, be much larger, since all the $\Delta m = \pm 1$ transitions are resonant at the same frequency and, if saturated, the RF will precess the full polarization giving the full amplitude equal to the off-resonance asymmetry.

\section{Results and analysis \label{sec:results}}

\subsection{Spin-Lattice Relaxation \label{sec:results:relaxation}}

\begin{figure}
\centering
\includegraphics[width=1.0\columnwidth]{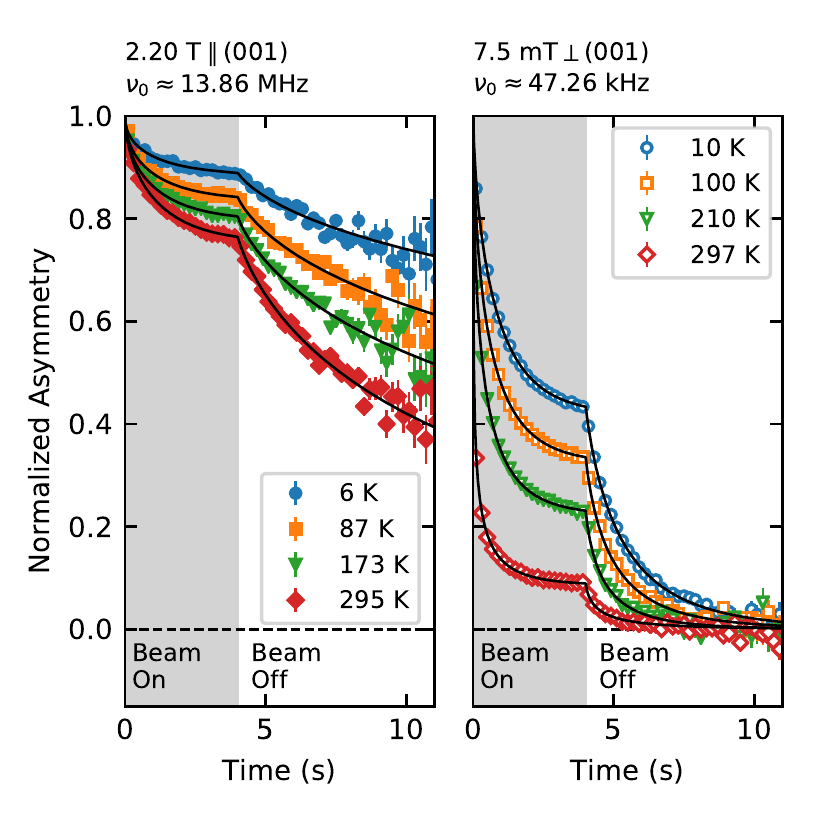}
\caption{ \label{fig:bts-slr-spectra}
\ce{^{8}Li} spin-lattice relaxation data in \ce{Bi2Te2Se} at high and low magnetic fields.
The characteristic kink at $t=\SI{4}{\second}$ corresponds to the trailing edge of the \ce{^{8}Li^{+}} beam pulse.
The \ce{^{8}Li} relaxation is strongly field-dependent, increasing with decreasing magnetic field, and increases non-monotonically with increasing temperature.
The solid black lines are the result of a global fit to all spectra at a given field to a stretched exponential relaxation function, \Cref{eq:strexp}, convoluted with the square beam pulse~\cite{2006-Salman-PRL-96-147601, 2015-MacFarlane-PRB-92-064409}.
}
\end{figure}

Typical \ce{^{8}Li} \slr\ data at high and low magnetic field in BTS are shown in \Cref{fig:bts-slr-spectra} for several temperatures, where the spectra have been normalized by their $t=0$ asymmetry ($A_{0}$).
It is immediately evident that the \slr\ rates are strongly dependent on both temperature and field.
In high fields, the relaxation is relatively slow, but comparable to that observed in some elemental metals~\cite{2015-MacFarlane-SSNMR-68-1} and semimetals~\cite{2014-MacFarlane-PRB-90-214422}.
Surprisingly, the relaxation is \emph{faster} than in the normal state of the structurally similar \ce{NbSe2}~\cite{2006-Wang-PB-374-239}, despite a very much smaller carrier density.
However, it is also slower than in the 3D TI \ce{Bi_{1-x}Sb_{x}}~\cite{2014-MacFarlane-PRB-90-214422}.
At all temperatures, the relaxation rate increases monotonically with decreasing magnetic field.
Even at very low temperatures near \SI{\sim 10}{\kelvin}, where most excitations are frozen out, the \slr\ remains substantial.
The relaxation rate at low field is orders of magnitude faster than in Tesla fields, suggesting the importance of low field relaxation from the host lattice nuclear moments~\cite{2009-Hossain-PRB-79-144518, 2018-MacFarlane-JPSCP-21-011020}.
On top of the field dependence, there is also a strong temperature dependence;
the relaxation rate increases with increasing $T$, but this trend is non-monotonic and at least one temperature exists (per field) where the rate is \emph{maximized}.

We now consider a detailed analysis to quantify these observations.
First, we remark that the relaxation is non-exponential at \emph{all} temperatures and fields.
The precise origin for this remains unclear.
While it is well-known that magnetic relaxation of quadrupolar spins is multiexponential, this is not the case here, as the initial polarization following optical pumping is purely dipolar, yielding single exponential magnetic relaxation, similar to the conventional NMR case where all the quadrupole satellites are simultaneously saturated~\cite{2000-MacFarlane-PRL-85-1108}.
If, however, the relaxation is quadrupolar (due to a fluctuating EFG), then in the slow limit for $I=2$, the relaxation is biexponential~\cite{1982-Becker-ZNA-37-697}.
In light of this, we use a phenomenological stretched exponential, consistent with the approach adopted in conventional NMR in similar materials~\cite{2013-Nisson-PRB-87-195202,2013-Koumoulis-PRL-110-026602, 2014-Koumoulis-AFM-24-1519, 2016-Levin-JPCC-120-25196}.
Explicitly, for a \ce{^{8}Li^{+}} ion implanted at time $t^{\prime}$, the spin polarization at time $t>t^{\prime}$ follows: 
\begin{equation} \label{eq:strexp}
   R \left ( t, t^{\prime} \right ) = \exp \left \{ - \left  [ \lambda \left ( t-t^{\prime} \right ) \right ]^{\beta} \right \},
\end{equation}
where $\lambda \equiv 1/T_{1}$ is the relaxation rate and $0 < \beta \leq 1$ is the stretching exponent.
We find this to be the simplest model that fits the data well (without overparamaterizing it) across all temperatures and fields.
Using \Cref{eq:strexp} convoluted with the \SI{4}{\second} beam pulse~\cite{2006-Salman-PRL-96-147601, 2015-MacFarlane-PRB-92-064409}, \slr\ spectra grouped by magnetic field $B_{0}$ and implantation energy $E$ were fit simultaneously with a shared common initial asymmetry $A_{0}(B_0,E)$.
Note the statistical uncertainties in the data are strongly time-dependent and accounting for this is crucial in the analysis.
To find the global least-squares fit, we used custom \texttt{C++} code leveraging the \texttt{MINUIT}~\cite{minuit} minimization routines implemented within \texttt{ROOT}~\cite{root}.
The fit quality is good in each case ($\tilde{\chi}_{\mathrm{global}}^{2} \approx 1.02$) and a subset of the fit results are shown in \Cref{fig:bts-slr-spectra} as solid black lines.
The large values of $A_{0}$ extracted from the fits (\SI{\sim 10}{\percent} for $B_{0} \geq \SI{2.20}{\tesla}$ and \SI{\sim 16}{\percent} for $B_{0} \leq \SI{20}{\milli\tesla}$) are consistent with the full beam polarization, implying that there is no appreciable missing fraction due to a very fast relaxing component.
$A_{0}$ extracted from these fits are used to normalize the spectra in \Cref{fig:bts-slr-spectra}.
For all the fits, the stretching exponent $\beta \approx 0.5$, with a weak temperature dependence: in high field, decreasing slightly at temperatures below \SI{\sim 200}{\kelvin}.

\begin{figure}
\centering
\includegraphics[width=1.0\columnwidth]{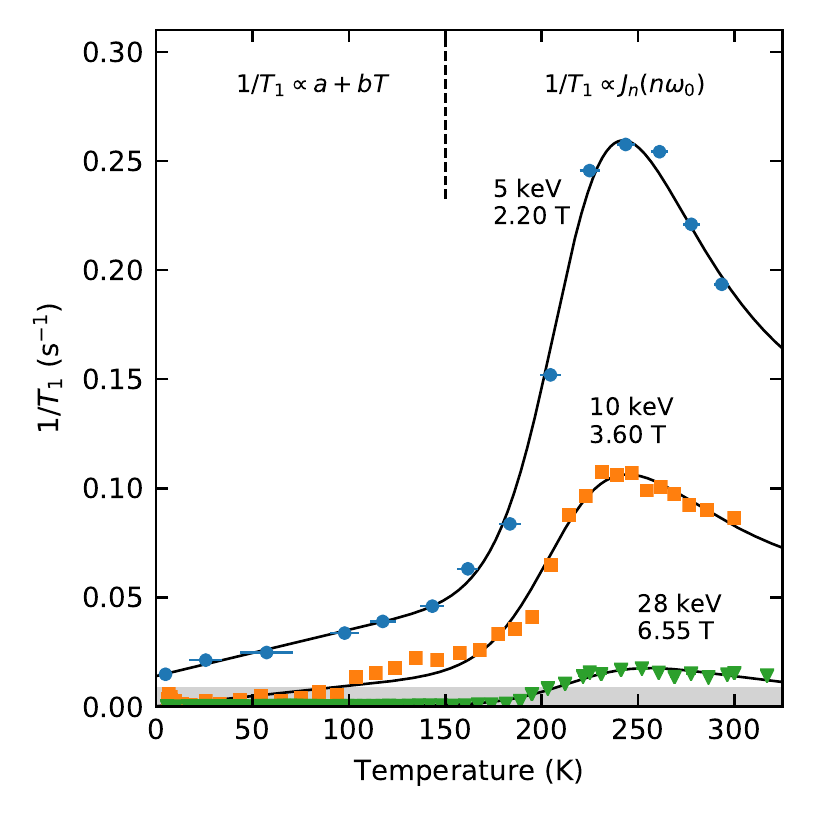}
\caption{ \label{fig:bts-slrfit-hf}
The spin-lattice relaxation rate for \ce{^{8}Li} in \ce{Bi2Te2Se} at high magnetic fields with $B_{0} \parallel (001)$ using a stretched exponential analysis. $1/T_{1}$ is strongly field-dependent, increasing with decreasing magnetic field, but also increases non-monotonically with temperature.
At each field, a clear $1/T_{1}$ maximum can be identified where the average fluctuation rate of the dynamics inducing relaxation matches the \ce{^{8}Li} Larmor frequency.
The solid black lines are fits to the model in \Cref{eq:rlx,eq:spectraldensity,eq:arrhenius}, consisting of a linear $T$-dependence with a non-zero intercept and a term due to ionic diffusion.
The highlighted grey region denotes the lower limit of measurable $1/T_{1}$ due to the \ce{^{8}Li} probe lifetime.
}
\end{figure}

The resulting values of $1/T_1$ are shown in \Cref{fig:bts-slrfit-hf}. 
Consistent with the qualitative behavior of the data, $1/T_1$ is relatively slow at high fields, but both temperature and field dependent.
Although the data also vary in implantation energy, the differences are primarily due to the applied field.
At the highest field, \SI{6.55}{\tesla}, the relaxation is extremely slow and near the measurable limit imposed by the \ce{^{8}Li} lifetime (grey region in \Cref{fig:bts-slrfit-hf}).
As the field is lowered, $1/T_{1}$ increases, and its temperature dependence is clearly non-monotonic.
Below \SI{\sim 150}{\kelvin}, $1/T_{1}$ increases approximately linearly with $T$, with a field dependent slope and $T \to 0$ intercept.
At higher temperatures, pronounced $1/T_{1}$ peaks are observed, whose maxima positions $T_{\mathrm{max}}$ are field dependent.
As $B_{0}$ is lowered, $T_{\mathrm{max}}$ shifts to slightly lower temperatures, with a more dramatic change when the field is decreased to \si{\milli\tesla} (see \Cref{fig:bts-slrfit-lf}).

\subsection{Modelling $T_1(T, \omega_{0})$ \label{sec:results:tone}}

\begin{figure}
\centering
\includegraphics[width=1.0\columnwidth]{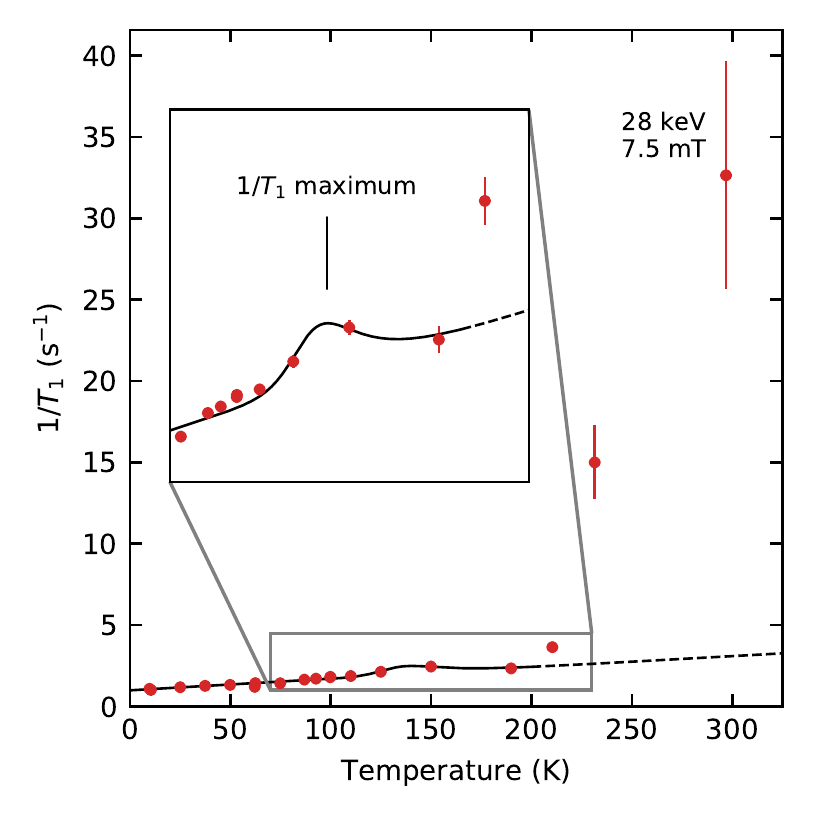}
\caption{ \label{fig:bts-slrfit-lf}
Temperature dependence of the \ce{^{8}Li^{+}} $1/T_{1}$ in \ce{Bi2Te2Se} at low magnetic field [$\SI{7.5}{\milli\tesla} \perp (001)$].
The relaxation rate is orders of magnitude larger than at high fields, with both a larger apparent $T$-linear slope and intercept.
There is some evidence of a small $1/T_{1}$ peak (see inset) near \SI{\sim 150}{\kelvin} superposed on the dominant linear dependence.
Above \SI{200}{\kelvin}, $1/T_{1}$ increases substantially, suggestive of another relaxation mechanism absent at higher fields. 
}
\end{figure}

We now consider a model of the temperature dependence of the measured $1/T_{1}$.
We interpret the relaxation peak as a Bloembergen-Purcell-Pound (BPP) peak~\cite{1948-Bloembergen-PR-73-679}, where the rate of a fluctuating interaction with the nuclear spin sweeps through the Larmor frequency $\omega_{0} = \gamma B_{0}$ at the rate peak~\cite{1948-Bloembergen-PR-73-679, 1979-Richards-TCP-15-141, 1988-Beckmann-PR-171-85}.
As we discuss in more detail below, we attribute this fluctuation to diffusive motion of interstitial \ce{^{8}Li^{+}} in the vdW gap between quintuple layers in BTS.
It is clear that the peaked relaxation adds to an approximately linearly temperature dependent term, reminiscent of Korringa relaxation characteristic of NMR in metals~\cite{1950-Korringa-P-16-601, 1990-Slichter-PMR}, and probably of electronic origin.
Based on this, we use the following model:
\begin{equation} \label{eq:rlx}
   1/T_{1} = a + b T + c \left ( J_{1} + 4J_{2} \right ).
\end{equation}
The first two terms in \Cref{eq:rlx} account for the linear $T$-dependence $\lambda_e \equiv a + bT$ and the remaining terms $\lambda_{\mathrm{diff}} \equiv c \left ( J_{1} + 4J_{2} \right )$ describe the peak.
\ldiff\ consists of a coupling constant, $c$, proportional to the mean-squared transverse fluctuating field and the $n$-quantum NMR spectral density functions, $J_{n}$~\cite{1988-Beckmann-PR-171-85}.
In this context, $J_{n}$ is a frequency dependent function peaked at $T_{\mathrm{max}}$, which occurs when the fluctuation rate driving relaxation matches $\sim n \omega_{0}$.
While the choice of a precise form of $J_{n}$ depends in detail on the dynamics, we use the empirical expression of Richards~\cite{1979-Richards-TCP-15-141}, which gives the correct asymptotic limits for relaxation produced by 2D fluctations~\cite{1981-Sholl-JPCSSP-14-447}, originating, for example, from diffusion of \ce{^{8}Li^{+}} confined to the vdW gap.
Explicitly~\cite{1979-Richards-TCP-15-141},
\begin{equation} \label{eq:spectraldensity}
   J_{n} \approx \tau_{c} \ln \left [ 1 + \left ( n\omega_{0}\tau_{c} \right )^{-2} \right ],
\end{equation}
where $\omega_{0}$ is the Larmor frequency, and $\tau_{c}$ is the correlation time, assumed to follow an Arrhenius temperature dependence:
\begin{equation} \label{eq:arrhenius}
   \tau_{c}^{-1} = \tau_{0}^{-1} \exp \left [ -E_{A} / \left ( k_{B} T \right ) \right ],
\end{equation}
where $\tau_{0}^{-1}$ and $E_{A}$ are the prefactor and activation energy, respectively, with $T$ and $k_{B}$ retaining their usual meanings of the absolute temperature and Boltzmann constant.

\begin{table}
\centering
\caption{ \label{tab:bts-8li-hop-rate}
Results from the analysis of the $1/T_{1}$ relaxation peaks in \Cref{fig:bts-slrfit-hf,fig:bts-slrfit-lf} using \Cref{eq:rlx,eq:spectraldensity,eq:arrhenius}.
Here, $\measuredangle$ denotes the orientation of the \ce{Bi2Te2Se} trigonal $c$-axis with respect to the applied field $B_{0}$ and $E$ is the \ce{^{8}Li^{+}} implantation energy.
Values for the coupling constant $c$, prefactor $\tau_{0}^{-1}$, and activation energy $E_{A}$ are indicated.
For comparison, the kinetic parameters extracted from fitting $\omega_{0}(T_{\mathrm{max}})$ to \Cref{eq:arrhenius} in \Cref{fig:bts-8li-hop-rate} (described in \Cref{sec:discussion:diffusion}) are shown in the bottom row.
The good agreement in these values, independent of the analysis details, indicates a single common dynamic process.
Differences in the parameter pairs $\tau_{0}^{-1}$ and $E_{A}$ may be attributed to the empirical Meyer-Neldel rule~\cite{2006-Yelon-RPP-69-1145}.
}
\begin{tabular}{SSSSSS}
\toprule
{$B_{0}$ (\si{\tesla})}	&	{$\measuredangle$}	&	{$E$ (\si{\kilo\electronvolt})}	&	{$c$ (\SI{e6}{\per\second\squared})}	&	{$\tau_{0}^{-1}$ (\SI{e12}{\per\second})}	&	{$E_{A}$ (\si{\electronvolt})} \\
\midrule
6.55	&	$\parallel$	&	28	&	1.88 \pm 0.02	&	1.02 \pm 0.13	&	0.156 \pm 0.003	\\
3.60	&	$\parallel$	&	10	&	5.08 \pm 0.10	&	0.7 \pm 0.2	&	0.153 \pm 0.006	\\
2.20	&	$\parallel$	&	20	&	2.86 \pm 0.04	&	28 \pm 7	&	0.235 \pm 0.005	\\
2.20	&	$\parallel$	&	5	&	7.23 \pm 0.16	&	1.2 \pm 0.5 &	0.173 \pm 0.009	\\
0.0075	&	$\perp$	   &	28	&	0.07 \pm 0.01	&	1	&	0.164 \pm 0.003	\\
\midrule
         &	            &	   &	               &	0.8 \pm 0.3	&	0.185 \pm 0.008	\\
\bottomrule
\end{tabular}
\end{table}

\begin{figure}
\centering
\includegraphics[width=1.0\columnwidth]{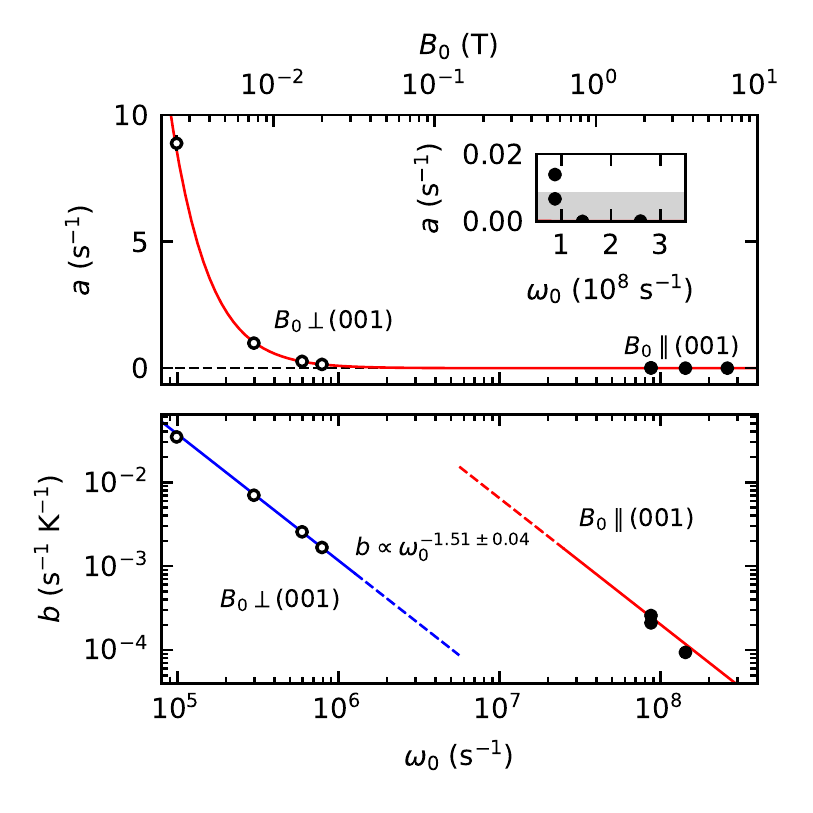}
\caption{ \label{fig:bts-slrfit-intercept-slope}
Field dependence of the intercept $a$ and slope $b$ describing the low temperature electronic relaxation in \Cref{eq:rlx}.
(Top panel): The $T \rightarrow 0$ intercept of $1/T_{1}$ falls with increasing field as $1/\omega_0^2$, behavior consistent with a low field relaxation due to fluctuating host lattice nuclear spins~\cite{2009-Hossain-PRB-79-144518}.
The solid red line indicates a fit to a simple BPP model~\cite{1948-Bloembergen-PR-73-679}, yielding a coupling constant $\bar{C}_{0}^{2} = \SI{1.5 \pm 0.3 e6}{\per\second\squared}$ and correlation time $\tau_{c} = \SI{3.2 \pm 0.7 e-5}{\per\second}$, the latter being remarkably close to the \ce{^{209}Bi} NMR $T_{2}$ time in \ce{Bi2Se3}~\cite{2012-Young-PRB-86-075137}.
The inset shows a zoom of the small high field intercepts, at the lower limit measurable due to the \ce{^{8}Li} probe lifetime highlighted in grey.
(Bottom panel): Field dependence of the low temperature slope of $1/T_{1}$, exhibiting a strong orientation dependence, but a common field dependence $\propto \omega_{0}^{-1.5}$, indicated by the solid coloured lines.
}
\end{figure}

Fits of the $1/T_{1}$ data to the model given by \Cref{eq:rlx} with \Cref{eq:spectraldensity,eq:arrhenius} are shown in \Cref{fig:bts-slrfit-hf,fig:bts-slrfit-lf} as solid black lines, clearly capturing the main features.
While the kinetic parameters determining the position and shape of the SLR peaks differed somewhat at different fields (see \Cref{tab:bts-8li-hop-rate}), good overall agreement was found from this analysis, and we consider the details further below in \Cref{sec:discussion:diffusion}.
As anticipated, the slope and intercept of $\lambda_e$ were both strongly field dependent.
At low fields, the intercept $a$ is quite large and varies as $\omega_{0}^{-2}$ (\Cref{fig:bts-slrfit-intercept-slope}), behavior characteristic of relaxation due to host nuclear spins.
Just as in the case of \ce{^{8}Li} in \ce{NbSe2}~\cite{2009-Hossain-PRB-79-144518}, this field dependence can be described by a simple BPP model~\cite{1948-Bloembergen-PR-73-679}: $1/T_{1} \propto \bar{C}_{0}^{2} \left [ 2\tau_{c}/ ( 1 + \omega_{0}^{2}\tau_{c}^{2}) \right ]$, where $\bar{C}_{0}^{2}$ is a coupling constant related to the fluctuating field from host nuclear dipoles.
A fit to this model, shown as a solid red line in the upper panel in \Cref{fig:bts-slrfit-intercept-slope}, yields $\bar{C}_{0}^{2} = \SI{1.5 \pm 0.3 e6}{\per\second\squared}$ and $\tau_{c} = \SI{3.2 \pm 0.7 e-5}{\second}$.
We note that the magnitude of $\bar{C}_{0}^{2}$ is compatible with the observed \ce{^{8}Li} resonance linewidth, discussed in \Cref{sec:results:resonance}, and that $\tau_{c}$ is remarkably close to the \ce{^{209}Bi} NMR $T_{2}$ time in \ce{Bi2Se3}~\cite{2012-Young-PRB-86-075137}.
It is, however, not clear whether the intercept at \SI{2.20}{\tesla} is a remnant of this low field relaxation, or is due to an electronic mechanism~\cite{1986-Gan-PRB-33-3595}.
In any case, it is small and we will not consider it further.
In contrast, the slope $b$ varies as $\omega_{0}^{-1.5}$ with a pronounced anisotropy between orientations (\Cref{fig:bts-slrfit-intercept-slope}).
We return to these points below in \Cref{sec:discussion:electronic}.

\subsection{Resonance Spectra \label{sec:results:resonance}}

\begin{figure}
\centering
\includegraphics[width=1.0\columnwidth]{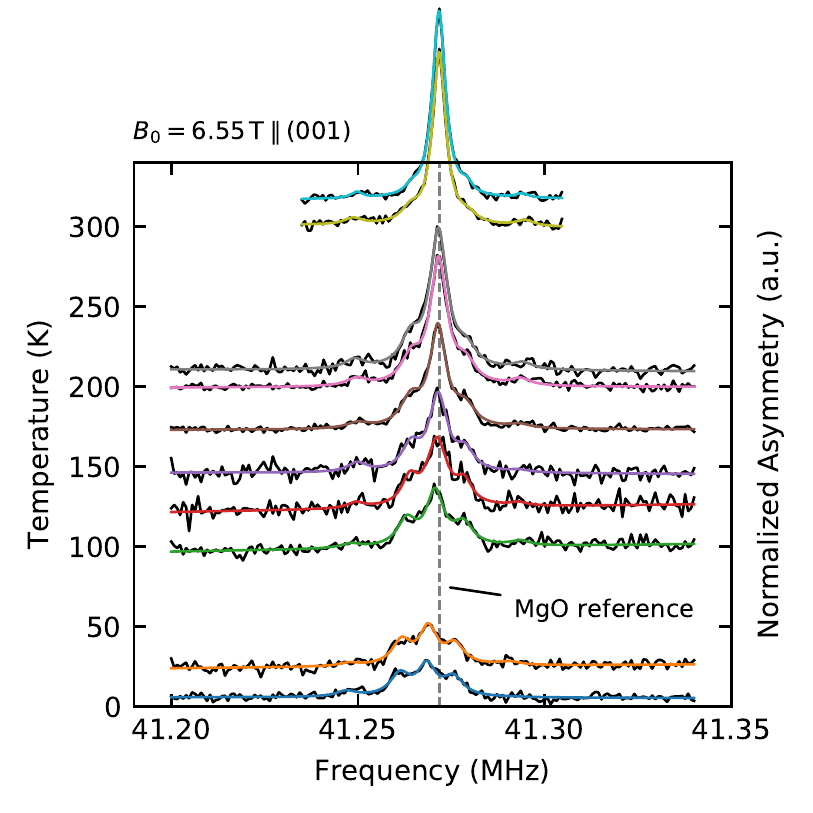}
\caption{ \label{fig:bts-1f-spectra}
\ce{^{8}Li} NMR spectra in \ce{Bi2Te2Se} with $\SI{6.55}{\tesla} \parallel (001)$ shown as solid black lines.
The vertical scale is the same for all the spectra, but the baselines are offset to match the absolute temperature indicated on the left.
The spectra reveal a small, temperature-independent quadrupole splitting, centred about an unsplit Lorentzian line, whose amplitude grows above \SI{100}{\kelvin}, becoming the dominant feature at higher temperatures.
The solid coloured lines are global fits to a sum of this Lorentzian plus $2I=4$ quadrupolar satellites of the quadrupole split resonance (see text for further details).
The reference frequency, determined by the \ce{^{8}Li} resonance position in \ce{MgO} (100) at room temperature, is indicated by the vertical dashed grey line.
Note the small negative shift of the \ce{Bi2Te2Se} resonances as the temperature is lowered.
}
\end{figure}

We now turn to the \ce{^{8}Li} resonance in BTS.
Typical spectra at high field are shown in \Cref{fig:bts-1f-spectra}.
Consistent with a non-cubic crystal, the \ce{^{8}Li} NMR is quadrupole split (as described in \Cref{sec:experiment:technique}).
This is confirmed by the helicity-resolved spectra (see \Cref{fig:bts-1f-helicities} in \Cref{sec:helicities})
which show 
opposite satellites in opposite helicities~\cite{2015-MacFarlane-SSNMR-68-1}. The EFG that produces the splitting is characteristic of the \eli\ site in the crystal.
From the spectra, it is relatively small, on the order of a few \si{\kilo\hertz}.
In addition, a second unsplit line is evident near the midpoint of the quadrupole pattern.
This line must be due to a distinct \eli\ site with a small, unresolved quadrupolar splitting.
A similar unsplit resonance was observed in the vdW layered material \ce{NbSe2}~\cite{2006-Wang-PB-374-239}, which was thought to originate from implanted \ce{^{8}Li^{+}} stopped at an interstitial site within the vdW gap.

This evidence for two sites suggests that our single component relaxation model in \Cref{sec:results:tone} is too simple.
However, more complicated relaxation functions suffer from overparametrization, and we have retained the single stretched exponential to represent the overall average relaxation of \eli\ in BTS.

The resonance was found to evolve substantially with temperature, as shown in \Cref{fig:bts-1f-spectra}.
Here, the spectra have been normalized to the off-resonance steady-state asymmetry to account for the variation of intensity due to SLR~\cite{2009-Hossain-PB-404-914}.
While it is apparent that the satellite intensities and splitting remain nearly temperature independent, the amplitude of the central line increases significantly above \SI{\sim 100}{\kelvin}, becoming the dominant feature by room temperature.
Additionally, a small shift in the resonance centre-of-mass frequency can be seen, increasing in magnitude with decreasing temperature.

To quantify these observations, we now consider a detailed analysis, noting first that the scale of the quadrupolar splitting is given by the quadrupole frequency~\cite{1957-Cohen-SSP-5-321}:
\begin{equation} \label{eq:nuq}
   \nu_q = \frac{e^{2} q Q}{8h}.
\end{equation}
As the splitting is small (\latin{i.e.}, $\nu_{q} \ll \nu_{0}$), the
satellite positions are given accurately by the first-order expression~\cite{1957-Cohen-SSP-5-321}:
\begin{equation} \label{eq:quadrupole}
   \nu_{i} = \nu_{0} - n_{i} \nu_{q} f ( \theta, \phi, \eta ) ,
\end{equation}
where $n_{i} = \pm 1 (\pm 3)$ for the inner (outer) satellites.
The angular factor
\begin{equation} \label{eq:angles}
   f ( \theta, \phi, \eta ) = \frac{1}{2} \left ( 3\cos^{2}\theta - 1 + \eta \sin^{2}\theta \cos 2\phi \right ),
\end{equation}
scales the splittings according to the polar ($\theta$) and azimuthal ($\phi$) angles between the external field and the EFG principal axis.
The parameter $\eta \in [0,1]$ is the EFG asymmetry, which for axially symmetric sites is zero.
From $\nu_{0}$, we additionally calculate the frequency shift, $\delta$, in parts per million (ppm) using:
\begin{equation} \label{eq:shift}
   \delta = 10^{6} \left ( \frac{\nu_{0} - \nu_{\ce{MgO}} }{ \nu_{\ce{MgO}} } \right ) ,
\end{equation}
where $\nu_{\ce{MgO}}$ is the reference frequency position in \ce{MgO} at \SI{300}{\kelvin} with $B_{0} \parallel (100)$~\cite{2014-MacFarlane-JPCS-551-012033}.

The helicity-resolved spectra were fit using the quadrupole splitting above with $\nu_0$ and $\nu_q$ as free parameters, in addition to line widths and amplitudes.
Similar to the \slr\ spectra in \Cref{sec:results:relaxation}, a global fitting procedure was used by way of \texttt{ROOT}'s~\cite{root} implementation of \texttt{MINUIT}~\cite{minuit}.
The two helicities of each spectrum were fit simultaneously with resonance positions and widths as shared parameters.
The fits were constrained such that the centre-of-mass frequency $\nu_{0}$ was shared between the unsplit Lorentzian and quadrupole satellites.
Any difference in centre of mass of the split and unsplit lines was too small to measure accurately, and the two lines shift in unison with temperature, as can be seen in \Cref{fig:bts-1f-spectra}.
In addition, we assume the EFG principal axis is along the $c$-axis and $\eta=0$, making the angular factor unity.
This assumption does not affect accurate extraction of the splitting frequency, but precludes unambiguous identification of the EFG tensor elements.
Based on a simple point charge model of the lattice (see \Cref{sec:discussion:site}), all reasonable interstitial \ce{^{8}Li^{+}} sites retain the 3-fold rotation axis of the hexagonal unit cell, supporting this simplification.
The change in satellite splittings from preliminary measurements at low field with $B_{0} \perp c$ are consistent with an EFG principal symmetry axis parallel to $c$.

\begin{figure}
\centering
\includegraphics[width=1.0\columnwidth]{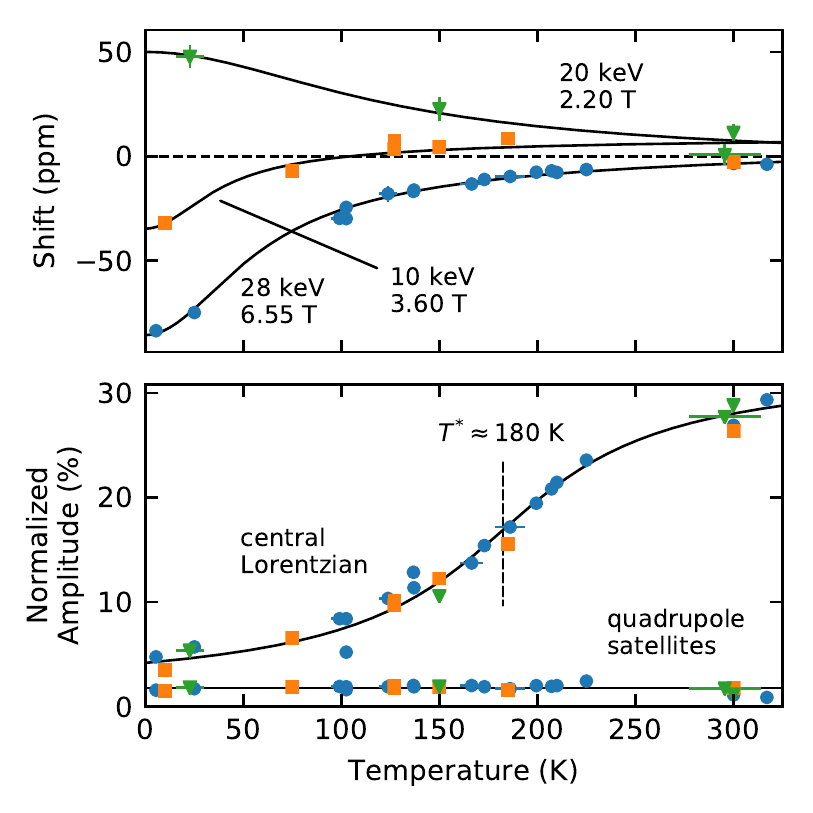}
\caption{ \label{fig:bts-1f-fits}
Fit results for \ce{^{8}Li} resonance in \ce{Bi2Te2Se} at high field with $B_{0} \parallel (001)$, following 
the fitting procedure described in the text.
The upper panel shows the resonance shift $\delta$ relative to \ce{MgO}.
Above \SI{\sim 150}{\kelvin}, the shift is nearly field independent and scattered about \num{0};
however, at lower temperatures, the trends at each field diverge, revealing a field dependent shift whose magnitude is maximized at the lowest temperatures.
The lower panel shows the change in amplitude of the central Lorentzian and quadrupole satellites (averaged over all four lines), normalized by the off-resonance baseline~\cite{2009-Hossain-PB-404-914}.
The amplitude of the unsplit line grows above \SI{\sim 100}{\kelvin}, contrasting the temperature independence of the quadrupole satellite amplitudes.
The inflection point $T^{*}$ of this trend is indicated by the dashed vertical black line.
The solid black lines are drawn to guide to the eye.
}
\end{figure}

The main parameters extracted from this analysis are shown in \Cref{fig:bts-1f-fits}.
Consistent with the two main qualitative features of the spectra, both the amplitude and shift show substantial changes with temperature.
In the top panel of \Cref{fig:bts-1f-fits}, above \SI{\sim 150}{\kelvin}, the shift is nearly field independent and centred around \SI{\sim 0}{ppm}.
Shifts of this magnitude are difficult to quantify accurately and the scatter in the values is of the same order as systematic variations in the line position, making differences of this order not very meaningful.
In contrast, at lower temperatures, the shift trends for each field diverge, revealing a significant field-dependent shift whose magnitude is maximized at the lowest temperature.

The changes in amplitude of the central Lorentzian and quadrupole satellites (averaged over all four lines), normalized by their off-resonance baseline~\cite{2009-Hossain-PB-404-914}, appear in the bottom panel of \Cref{fig:bts-1f-fits}.
Consistent with \Cref{fig:bts-1f-spectra}, the amplitude of the unsplit line grows above \SI{\sim 100}{\kelvin} and approaches saturation near room temperature, in contrast to the temperature insensitivity of the satellite amplitudes.
From the smooth growth of the unsplit line amplitude, we identify the inflection point $T^{*} \approx \SI{180}{\kelvin}$ of the trend, as indicated in \Cref{fig:bts-1f-fits}.
The other spectral parameters are nearly independent of temperature.
The line widths are about \SI{\sim 6}{\kilo\hertz}, with the central Lorentzian narrowing slightly with increasing $T$.
This narrowing is likely the cause of its increase in amplitude.
Similarly, $\nu_{q} \approx \SI{7.4}{\kilo\hertz}$ was characteristic of the splittings over the entire measured temperature range.

Based on these results at the lowest measured temperatures, where dynamic contributions to the resonance are absent, we estimate a 1:1 relative occupation for \ce{^{8}Li^{+}} in the two sites.
Note, however, that changes in amplitude, up to our highest measured temperature \SI{317}{\kelvin}, are inconsistent with a site change, where the growth of one amplitude is at the expense of the other~\cite{2004-Morris-PRL-93-157601}.

\section{Discussion \label{sec:discussion}}

With the main results presented in \Cref{sec:results}, the remaining discussion is organized as follows:
in \Cref{sec:discussion:diffusion}, we consider the dynamics causing the high temperature $1/T_{1}$ peaks,
while \Cref{sec:discussion:electronic} considers the electronic properties of BTS giving rise to relaxation and resonance shifts at low temperature.

\subsection{High temperature lithium-ion diffusion \label{sec:discussion:diffusion}}

The most likely source of the relaxation at high temperatures is diffusive motion of \ce{^{8}Li^{+}}.
While we cannot rule out some local stochastic motion within a cage, or motion of another species in the lattice, given the demonstrated ability to chemically insert \ce{Li} at room temperature in isostructural bismuth chalcogenides~\cite{1988-Paraskevopoulos-MSEB-1-147, 2010-Bludska-JSSC-183-2813}, we expect a low barrier to interstitial diffusion for any implanted \ce{^{8}Li^{+}} in the vdW gap.
Stochastic motion causes the local magnetic field and EFG to become time dependent causing 
relaxation~\cite{1948-Bloembergen-PR-73-679, 1979-Richards-TCP-15-141}.
From the \slr, we can thus obtain information on the kinetic parameters of the diffusion.
From the model of $1/T_{1}(T)$ introduced in \Cref{eq:rlx} from \Cref{sec:results:tone}, we obtain the kinetic parameters listed in \Cref{tab:bts-8li-hop-rate}.
The success of the model is demonstrated by the good self-agreement in the barrier $E_{A}$ and prefactor $\tau_{0}^{-1}$, indicating that an activated dynamic process with a 2D spectral density provides a single common source for the observed \ldiff.
Thus, the correlation rate $\tau_{c}^{-1}$ in \Cref{eq:arrhenius} represents the atomic hop rate $\tau^{-1}$.
The small barrier, on the order of \SI{\sim 0.2}{\electronvolt}, is consistent with expectations for an isolated interstitial ion, where Coulomb \ce{Li^{+}}-\ce{Li^{+}} repulsion is negligible.
Similarly, the $\tau_{0}^{-1}$s on the order of \SI{\sim e12}{\per\second} are compatible with typical optical phonon frequencies, as is often the case for mobile ions in a lattice.
The relatively small deviations in $E_{A}$ and $\tau_{0}^{-1}$ obtained at different fields may be ascribed to the empirical Meyer-Neldel rule, where $\tau_{0}^{-1}$ increases exponentially with increasing $E_{A}$, as is often observed for related kinetic processes~\cite{2006-Yelon-RPP-69-1145}.

In general, the exponent appearing in $J_{n}$ from \Cref{eq:spectraldensity} can vary from \numrange{-1}{-2}~\cite{1994-Kuchler-SSI-70-434}, with deviations from \num{-2} reflecting correlated dynamics that can arise from, for example, Coulomb interactions with other ions acting to bias the probe ion's trajectory.
Such correlations affect the shape the $1/T_{1}$ peak, yielding a characteristic asymmetry with a shallower slope on the low-$T$ side.
In contrast, the high symmetry about $T_{\mathrm{max}}$ is consistent with uncorrelated fluctuations driving relaxation, as expected for isolated \ce{^{8}Li} undergoing direct interstitial site-to-site hopping.

\begin{figure}
\centering
\includegraphics[width=1.0\columnwidth]{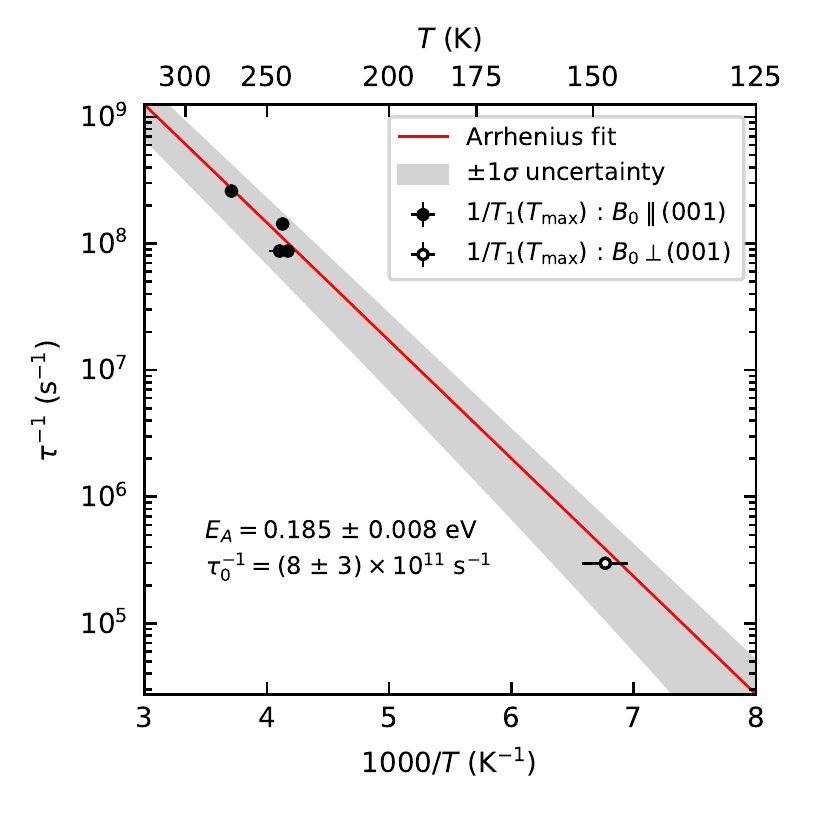}
\caption{ \label{fig:bts-8li-hop-rate}
Arrhenius plot of the \ce{^{8}Li} fluctuation rate extracted from the NMR frequency dependent positions of $1/T_{1}(T_{\mathrm{max}})$ in \Cref{fig:bts-slrfit-hf,fig:bts-slrfit-lf}.
The solid red line is a fit to \Cref{eq:arrhenius} with the activation energy $E_{A}$ and prefactor $\tau_{0}^{-1}$ indicated in the inset and the fit's $\pm 1 \sigma$ uncertainty band is highlighted in grey.
The kinetic parameters extracted from this minimally model-dependent analysis are in good agreement with those obtained from fits of the $1/T_{1}$ data to \Cref{eq:rlx,eq:spectraldensity,eq:arrhenius}, given in \Cref{tab:bts-8li-hop-rate}.
}
\end{figure}

As further confirmation of the appropriateness of the form of $J_{n}$, we consider an alternative approach agnostic to these details.
At each field, we determine the temperature $T_{\mathrm{max}}$ of the $1/T_{1}$ peak using a simple parabolic fit (after removal of the $T$-linear contribution).
This approach has the advantage that it does not rely on any particular form of $J_{n}$, and we recently used it to quantify diffusion of isolated \ce{^{8}Li^{+}} in rutile \ce{TiO2}~\cite{2017-McFadden-CM-29-10187}.
Finally, for each $T_{\mathrm{max}}$ we assume $\tau_{c}^{-1}$ matches the Larmor frequency $\omega_{0}$.
The results are shown in the Arrhenius plot in \Cref{fig:bts-8li-hop-rate}, where the linearity of the data, spanning three orders of magnitude, demonstrates the consistency of the approach.
The Arrhenius fit shown yielded an activation energy $E_{A} = \SI{0.185 \pm 0.008}{\electronvolt}$ and prefactor $\tau_{0}^{-1} = \SI{8 \pm 3 e11}{\per\second}$, in good agreement with the values from the analysis using the 2D $J_{n}$.
Noting that this result lies in the middle of range reported in \Cref{tab:bts-8li-hop-rate}, we take it as the best determination of the \ce{^{8}Li^{+}} hop rate.

Another well-known signature of diffusion in NMR is motional narrowing.
When the diffusive correlation rate exceeds the characteristic static frequency width of the line, the local broadening interactions are averaged and the line narrows. In the context of dilute interstitial diffusion in a lattice, the primary quadrupolar interaction may, however, not be averaged to zero, since, in the simplest case, each site is equivalent and characterized by the same EFG.
We observe a slight narrowing and a large enhancement in the amplitude of the unsplit resonance with an onset in the range \SIrange{100}{120}{\kelvin}, consistent with where the extrapolated $\tau_{c}^{-1}$ would be in the \si{\kilo\hertz} range of the linewidth.
With the \cw\ resonance measurement, we often find the change in amplitude is more pronounced than the width~\cite{2011-Saadaoui-PRB-83-054504}.

\begin{figure}
\centering
\includegraphics[width=1.0\columnwidth]{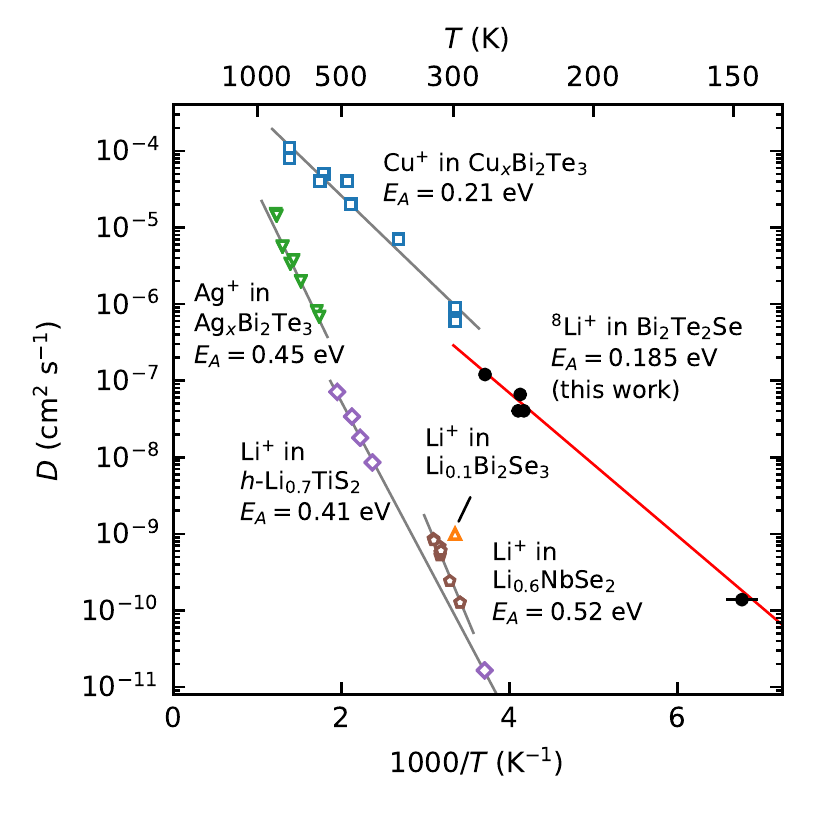}
\caption{ \label{fig:diffusion-comparison}
Arrhenius plot of the diffusion coefficient $D$ for \ce{^{8}Li^{+}} in \ce{Bi2Te2Se}.
Values for $D$, shown in black, were estimated using the Einstein-Smoluchouski expression [\Cref{eq:einstein-smoluchouski}], taking $f=1$, $d=2$, and $l = \SI{4.307}{\angstrom}$, with the values of $\tau^{-1}$ taken from \Cref{fig:bts-8li-hop-rate}.
For comparison, diffusion coefficients for \ce{Li^{+}}~\cite{1986-Folinsbee-MRB-21-961, 1988-Paraskevopoulos-MSEB-1-147, 2006-Wilkening-PRL-97-065901}, \ce{Ag^{+}}~\cite{1963-Keys-JPCS-24-563}, and \ce{Cu^{+}}~\cite{1960-Carlson-JPCS-13-65, 1992-Maclachlan-JMS-27-4223} in structurally related materials are also included.
It its clear that the diffusivity of isolated \ce{^{8}Li^{+}}, free from repulsive \ce{Li^{+}}-\ce{Li^{+}} interactions, is exceptional, significantly exceeding that in the well-known fast ion conductor $h$-\ce{Li_{x}TiS2}~\cite{2006-Wilkening-PRL-97-065901, 2008-Wilkening-PRB-77-024311}.
}
\end{figure}

Using the \ce{^{8}Li^{+}} hop rate from above, we convert $\tau^{-1}$ to diffusivity via the Einstein-Smolouchouski expression:
\begin{equation} \label{eq:einstein-smoluchouski}
   D = f\frac{l^{2}}{2d\tau},
\end{equation}
where $l$ is the jump length, $d=2$ is the dimensionality, and $f$ is the correlation factor, assumed to be unity for direct interstitial diffusion, to compare with other measurements of interstitial ionic diffusion in related materials.
Using $l\approx\SI{4.307}{\angstrom}$, the distance between neighbouring $3b$ sites in the vdW gap in the ideal BTS lattice (see \Cref{fig:bts-vdwg-sites} in \Cref{sec:discussion:site}),
we estimate $D$ for \ce{^{8}Li^{+}}, finding a value on the order of \SI{e-7}{\centi\meter\squared\per\second} at \SI{300}{\kelvin}.
An Arrhenius plot comparing the diffusivity of \ce{^{8}Li^{+}} in BTS with other cations in structurally related materials~\cite{1960-Carlson-JPCS-13-65, 1963-Keys-JPCS-24-563, 1986-Folinsbee-MRB-21-961, 1988-Paraskevopoulos-MSEB-1-147, 1992-Maclachlan-JMS-27-4223, 2006-Wilkening-PRL-97-065901} is shown in \Cref{fig:diffusion-comparison}.
Note that most of these diffusion coefficients were determined using either radiotracer~\cite{1960-Carlson-JPCS-13-65, 1963-Keys-JPCS-24-563} or transient electrochemical~\cite{1986-Folinsbee-MRB-21-961, 1992-Maclachlan-JMS-27-4223, 1988-Paraskevopoulos-MSEB-1-147} techniques.
For $h$-\ce{Li_{0.7}TiS2}, we used the \ce{Li^{+}} hop rate determined by \ce{^{7}Li} NMR~\cite{2006-Wilkening-PRL-97-065901} and \Cref{eq:einstein-smoluchouski}, taking $f=1$, $d=2$, and $l=\SI{2.14}{\angstrom}$ (the distance between $1b$ and $2d$ sites)~\cite{2008-Wilkening-PRB-77-024311}.

It is clear that the mobility of isolated \ce{^{8}Li^{+}} is exceptional;
our estimate for $D$ greatly exceeds that of lithium in the well-known fast ion conductor $h$-\ce{Li_{x}TiS2}~\cite{2006-Wilkening-PRL-97-065901, 2008-Wilkening-PRB-77-024311}.
Similarly, the lithium diffusion coefficient in lithium intercalated \ce{Bi2Se3} is considerably slower~\cite{1988-Paraskevopoulos-MSEB-1-147}, possibly due to \ce{Li^{+}}-\ce{Li^{+}} interaction.
Interestingly, the mobility of \ce{Cu} in isostructural \ce{Bi2Te3} is also extremely high, as revealed by \ce{^{64}Cu} radiotracer~\cite{1960-Carlson-JPCS-13-65} and electrochemical methods~\cite{1992-Maclachlan-JMS-27-4223}.
Lastly, we note that similarly large $D$ values were reported recently for \ce{^{8}Li^{+}} in the one dimensional ion conductor rutile \ce{TiO2}~\cite{2017-McFadden-CM-29-10187} and we speculate that the exceptional mobility may be generic for \emph{isolated} \ce{Li^{+}} (\latin{i.e.}, at infinitely dilute concentrations) in ion conducting solids.
It would be interesting to test this conjecture against detailed \latin{ab initio} calculations.
Understanding the mobility of dilute intercalates, a simple theoretical situation difficult to interrogate experimentally, remains of fundamental interest~\cite{2016-Gosalvez-PRB-93-205416}.

\subsection{Low temperature electronic properties \label{sec:discussion:electronic}}

We turn now to the low temperature results, below the diffusion-related peak, in the range \SIrange{5}{150}{\kelvin}.
The diffusive contribution to the relaxation rate is falling exponentially with reduced temperature below the peak,
and the remaining low temperature relaxation must thus have a distinct origin. In this regime, 
the relaxation rate is linear in $T$, with a slope that is significantly field dependent.
In this range there is also a $T$-dependent resonance shift that is also field dependent.

In metals, the dominant features of NMR (and \bnmr) are due to coupling between the conduction electron and nuclear spins, giving rise to the Korringa $T$-linear $1/T_1$ and a $T$-independent Knight shift proportional to the Pauli susceptibility~\cite{1950-Korringa-P-16-601, 1990-Slichter-PMR}.
The apparent Korringa-like dependence in \Cref{fig:bts-slrfit-hf,fig:bts-slrfit-lf} is thus surprising, since ideal BTS is a narrow gap semiconductor, with an energy gap $E_g \approx \SI{0.3}{\electronvolt}$~\cite{2012-Akrap-PRB-86-235207}.
In comparison to metals, NMR in semiconductors is much less well-known, largely because the coupling to the electron spins is much less universally dominant, and often other channels, such as phonons, compete with electronic effects and complicate the interpretation (\latin{e.g.}, in elemental \ce{Te}~\cite{1979-Selbach-PRB-19-4435} and \ce{InSb}~\cite{1969-Bridges-PR-182-463}, with similar $E_g$ to BTS).
However, while calculations and experiments agree on $E_g$ in BTS~\cite{2012-Neupane-PRB-85-235406, 2012-Akrap-PRB-86-235207}, it is abundantly clear from both experiment~\cite{2012-Jia-PRB-86-165119} and theory~\cite{2011-Hashibon-PRB-84-144117, 2012-Scanlon-AM-24-2154} that it is unlikely to exist as an intrinsic semiconductor.
Rather, it is significantly self-doped by native defects, such as chalcogenide vacancies (donor) and \ce{Bi}/\ce{Te} antisite defects (acceptor).
Among the tetradymites, BTS is relatively highly insulating, but this does not indicate a paucity of native defects, but rather a coincidental near compensation between the $n$- and $p$-types.
\@ \citeauthor{2015-Brahlek-SSC-215-54} have made the point that these materials are, in fact, so highly doped that they are metallic, but are poor conductors due to disorder~\cite{2015-Brahlek-SSC-215-54}.
Such disorder is clearly evident in our data, contributing to the stretched exponential character of the spin relaxation and the line broadening.

The origin of the linearity of the Korringa law is evident from a derivation based on a Fermi golden rule approach to the spin-flip scattering of conduction electrons by the hyperfine interaction with the nuclear spin.
The sum over electron momenta when converted to an integral over energy yields:
\begin{equation} \label{eq:goldenrule}
  \frac{1}{T_1} = \frac{2\pi}{\hbar} A^2 \int \rho^2(E) f(E) \left [ 1-f(E) \right ] \, \mathrm{d}E,
\end{equation}
where $A$ is the hyperfine coupling energy, $f(E)$ is the Fermi-Dirac distribution, $\rho(E)$ is the electronic density of states, and the integral is over all electron energies in the conduction band. In a broad band degenerate metal, where the Fermi level $E_F \gg k_{B}T$, the density of states $\rho(E)$ is practically constant over the range where the Fermi factor is nonzero, and the integral is given to an excellent approximation by:
\begin{equation} \label{eq:korringa}
  \frac{1}{T_1} = \frac{2\pi}{\hbar} A^{2} \rho^{2}(E_F) k_{B}T.
\end{equation}
The Korringa slope is thus determined by the square of the product of $A$ and $\rho(E_F)$.  
In fact, on inspection of the \ce{^{8}Li} \lle\ in BTS, we find it comparable to wide band metals with vastly higher carrier densities~\cite{2004-Morris-PRL-93-157601, 2006-Wang-PB-374-239, 2007-Salman-PRB-75-073405, 2008-Parolin-PRB-77-214107, 2009-Hossain-PB-404-914, 2009-Wang-PB-404-920, 2012-Ofer-PRB-86-064419, 2018-MacFarlane-JPSCP-21-011020}.
While the coupling $A$ for implanted \elip\ in BTS is not known, it is unlikely to compensate for the much lower $\rho(E_F)$, to yield a comparable Korringa slope.
Note that the Korringa law is remarkably robust to disorder and, for example, applies in the normal state of the alkali fullerides where the mean free path is comparable to the lattice constant (the Ioffe-Regel limit)~\cite{1996-Pennington-RMP-68-855}.
However, in highly disordered metals, the slope is strongly enhanced~\cite{1971-Warren-PRB-3-3708, 1983-Gotze-ZPB-54-49, 1994-Shastry-PRL-72-1933}.
Such an enhancement may account for the substantial slope we observe.
However, \Cref{eq:goldenrule} indicates that if $\rho(E)$ has significant structure on scales comparable to $k_{B}T$, as it might in a narrow impurity band, its detailed form and nondegeneracy can render the $T$-dependence nonlinear~\cite{2005-Hoch-PRB-71-035115}, as found, for example, near the metal-insulator transition (MIT) in doped silicon~\cite{2005-Meintjes-PRB-71-035114}.
Thus, the evident linearity is still surprising.
Moreover, in metals, the Korringa slope does \emph{not} depend on magnetic field.

In doped \ce{Si}, near the MIT, a field dependence of the enhanced Korringa slope has been found at \si{\milli\kelvin} temperatures~\cite{1985-Paalanen-PRL-54-1295}, where it was attributed to the occurrence of uncompensated localized electron spins, probably on some subset of more isolated neutral \ce{P} donors.
Such moments are also evident in the NMR of the dopant nuclei~\cite{1987-Alloul-PRL-59-578}.
However, in BTS (in contrast to \ce{Si}), due to the high dielectric constant and low effective mass, \emph{magnetic carrier freezeout}~\cite{1956-Yafet-JPCS-1-137, 1969-Dyakonov-PR-180-813} may account for the diminished slope at high fields.
In this case, the field localizes the carriers so they no longer participate in the conduction band, correspondingly reducing the Korringa slope.
This may account, in part, for the significant positive magnetoresistance in BTS~\cite{2012-Xiong-PE-44-917, 2013-Assaf-APL-102-012102}.
Similarly, the field and temperature dependent shifts in \Cref{fig:bts-1f-fits} may reflect a constant (diamagnetic) contribution (\latin{cf.} the \ce{^{125}Te} shift~\cite{2014-Koumoulis-AFM-24-1519}), in addition to a positive hyperfine field related to the carriers, which diminishes with localization.
However, the localized electrons may well provide additional (more inhomogeneous) relaxation and resonance broadening.
We know of no case where NMR has been used to study magnetic freezeout. 
In order to test this idea, it will be essential to compare results on different samples of BTS and related materials (\latin{e.g.}, \ce{Ca}-doped \ce{Bi2Se3}).

Above, we only considered bulk origins for the \lle.
While we do not expect any direct coupling to the topological surface state at the implantation energies used, it is important to consider the effects of band bending at the surface.
If the bulk electronic bands are bent downward at the surface, a conventional 2D electron gas may be stabilized~\cite{2010-Bianchi-NC-1-128}.
Metallic screening will confine this surface region to a few \si{\nano\meter} from the surface.
In the opposite case, upward bending makes the surface region more insulating, and the dielectric screening is much weaker, producing a depletion region on the scale of \si{\micro\meter}, as has been demonstrated by ionic liquid gating~\cite{2013-Xiong-PRB-88-035128}.
In the latter case, the acceptor band is depopulated by the surface dipole.
Calculations suggest this is not the case~\cite{2015-Fregoso-JPCM-27-422001}, but there is substantial evidence for time-dependent band bending from ARPES~\cite{2017-Frantzeskakis-PRX-7-041041}.
Generally, this downward bending is found to produce a more metallic surface.
With its exposure to air prior to the measurements, we assume that our sample's surface is passivated and is metallic, typical of an ``aged'' surface, so we can safely neglect possible depth dependence to the carrier concentration at the implantation energies used.
This is consistent with the absence of an appreciable implantation energy dependence in \lle\ at \SI{2.20}{\tesla} (not shown).

We have, so far, focused exclusively on the interaction between \ce{^{8}Li} and the electron \emph{spins}.
We should also consider its interaction with their orbital currents.
Note that at \SI{100}{\kelvin} $1/T_1$ in BTS is comparable to semimetallic bismuth~\cite{2014-MacFarlane-PRB-90-214422}, with a similar carrier density, but a much longer mean free path.
In \ce{Bi}, due to its strong orbital diamagnetism, we suggested that orbital fluctuations might be responsible for the fast relaxation of \ce{^{8}Li} and its concentration dependence in \ce{Bi_{x}Sb_{1-x}} solid solutions~\cite{2014-MacFarlane-PRB-90-214422}.
This mechanism is due to fluctuating electronic currents, so it is naturally related to the conductivity $\sigma$.
Specifically~\cite{2018-Maebashi-JPCS}:
\begin{equation}
  \left( \frac{1}{T_1} \right)_{\mathrm{orb}} \propto k_{B}T \int \frac{ \mathrm{Re} \left \{ \sigma_\perp(q,\omega_{0}) \right \} }{q^2} \, \mathrm{d}^{3}q ,
\end{equation}
where $\sigma_\perp(q,\omega)$ is the generalized wavevector ($q$) and frequency dependent conductivity transverse to the nuclear spin.
Clearly, this relaxation is enhanced by higher conductivity;
however, in metals, it has long been recognized that orbital relaxation is usually much weaker than the spin-related Korringa relaxation discussed above.

In a layered conductor, the hyperfine coupling $A$ for interstitial \eli\ can be particularly weak (\latin{e.g.}, \ce{NbSe2}~\cite{2006-Wang-PB-374-239}).
In these circumstances, it is possible that orbital relaxation will dominate.
Unlike the contact coupling, orbital fields fall as $1/r^3$, where $r$ is the distance between the probe spin and the fluctuating current, so all nuclei in the material, and potentially even nuclei in close proximity, will sense spatially extended orbital fluctuations.
For example, this mechanism has been studied as a proximal source of decoherence in spin based qubit devices~\cite{2015-Kolkowitz-S-347-1129}.
\@ \citeauthor{1991-Lee-PRB-43-1223} have explicitly considered the case of orbital relaxation in a 2D layered metal~\cite{1991-Lee-PRB-43-1223}.
They find a weak logarithmic singularity in $1/T_1$ in the clean limit that is cut off by a finite mean free path~\cite{2007-Knigavko-PRB-75-134506}.
Similar to the Korringa rate, the orbital relaxation rate is linear in $T$ for a broad band metal.
Recently, this approach was generaralized to the case of massive Dirac-like electrons in 3D, appropriate to semimetallic \ce{Bi_{1-x}Sb_{x}}~\cite{2017-Hirosawa-JPSJ-86-063705, 2018-Maebashi-JPCS}.
They find $T$-linear relaxation when the chemical potential is outside the gap~\cite{2018-Maebashi-JPCS}, but within the gap, an anomalous relation that is explicitly field dependent via the NMR frequency $\omega_{0}$ is obtained, where $\lambda_{\mathrm{orb}} \propto T^{3} \ln(2 k_{B}T / \hbar \omega_0)$.
While this is not what we find, it does suggest that if orbital relaxation is effective here, it may exhibit some unexpected field dependence in the inhomogeneous metallic state hypothesized for the tetradymites~\cite{2017-Bomerich-PRB-96-075204}.
Similarly, orbital current fluctuations will almost certainly be accompanied by a fluctuating charge distribution (and hence EFG), giving rise to a quadrupolar relaxation in addition to the magnetic orbital relaxation. This has been considered explicitly for Dirac electrons in 3D, and it was concluded that the quadrupolar contribution was negligible~\cite{2018-Maebashi-JPCS}. While this result probably does not apply directly here, it suggests a theory for the relaxation in BTS could focus on a purely magnetic mechanism.

In fact, some features of our data do suggest the importance of orbital effects.
One is the similarity of shifts of the quadrupolar split resonance and the unsplit Lorentzian.
If, as seems to be reasonable, these resonances originate in different lattice sites of \ce{^{8}Li^{+}}, then one would expect different hyperfine couplings and different (spin) shifts.
If the shift is rather orbital in origin with significant contribution from long length-scale currents, one would expect the same shift for any site (and even any nucleus) in the unit cell.
In general, however, one would expect both spin and orbital couplings, and nuclei of such different species, such as \ce{^{209}Bi} or \ce{^{8}Li}, would likely differ.

We have ruled out a number of possibilities, but we do not have a conclusive explanation of the interesting features of the data at low temperatures.
At this point, it is worth noting that conventional NMR in the tetradymite TIs are also characterized by highly variable power law $T$-dependent relaxation~\cite{2012-Taylor-JPCC-116-17300, 2012-Young-PRB-86-075137, 2013-Nisson-PRB-87-195202} whose dependence on magnetic field has largely not been explored.

\section{Conclusion \label{sec:conclusion}}

Using temperature and field dependent ion-implanted \ce{^{8}Li} \bnmr, we studied the high temperature ionic and low temperature electronic properties of \ce{Bi2Te2Se}.
Two distinct thermal regions were found;
above \SI{\sim 150}{\kelvin}, the isolated \ce{^{8}Li^{+}} probe undergoes ionic diffusion with an activation energy $E_{A} = \SI{0.185 \pm 0.008}{\electronvolt}$ and attempt frequency $\tau_{0}^{-1} = \SI{8 \pm 3 e11}{\per\second}$ for atomic site-to-site hopping.
A comparison of the kinetic details with other well-known \ce{Li^{+}} conductors suggests an exceptional mobility of the isolated ion.
At lower temperature, field dependent relaxation and resonance shifts are observed.
While the linearity of the spin-lattice relaxation is reminiscent of a Korringa mechanism, existing theories are unable to account for the extent of the field dependence.
We suggest that these may be related to a strong contribution from orbital currents or the magnetic freezeout of charge carriers in the heavily compensated semiconductor.
Field dependent conventional NMR of the stable host nuclei, combined with the present data, will further elucidate their origin.

\begin{acknowledgments}
We thank: R.\ Abasalti, D.\ J.\ Arseneau, B.\ Hitti, S.\ Daviel, K.\ Olchanski, and D.\ Vyas for their excellent technical support;
D.\ E.\ Eldridge, Q.\ Song, and D.\ Wang for some help with the early measurements and analysis;
M.\ Ogata and J.\ A.\ Folk for useful discussions and comments;
K.\ Foyevtsova and P.\ Blaha for help with the density functional calculations;
and O.\ Prakash for assistance with transport measurements.
This work was supported by NSERC Discovery grants to R.F.K.\ and W.A.M., as well as NSERC CREATE IsoSiM Fellowships to R.M.L.M.\ and A.C.
The crystal growth at Princeton University was supported by the ARO-sponsored MURI on topological insulators, grant number W911NF1210461.
\end{acknowledgments}

\appendix

\section{Implantation profiles \label{sec:implantation}}

\begin{figure}
\centering
\includegraphics[width=1.0\columnwidth]{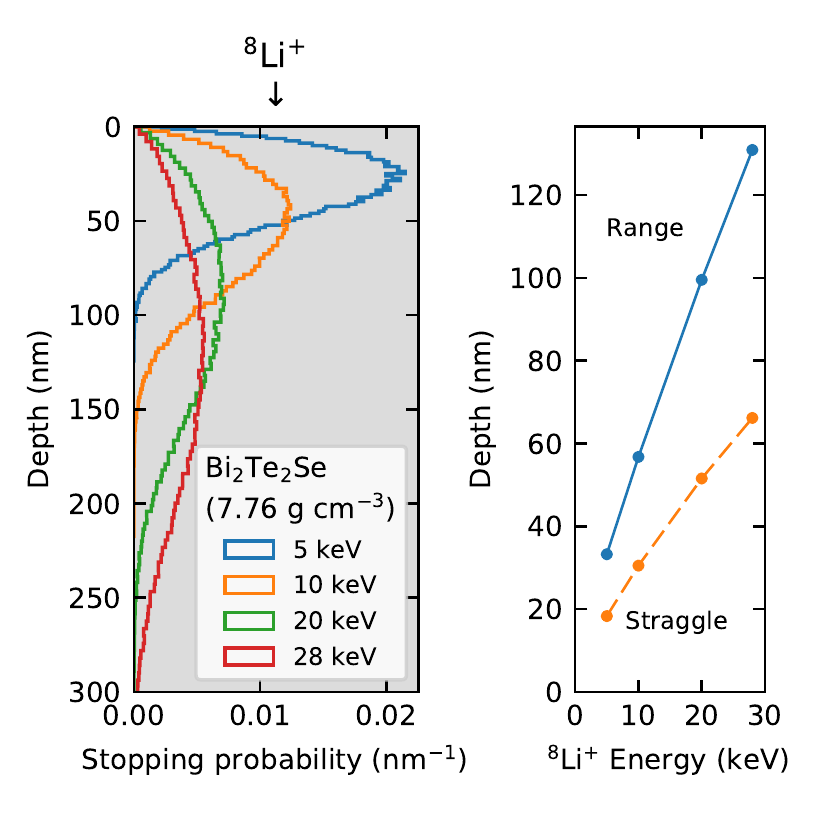}
\caption{ \label{fig:srim}
Stopping distribution and range for \ce{^{8}Li^{+}} implanted in \ce{Bi2Te2Se} calculated using the \texttt{SRIM} Monte Carlo code~\cite{srim}.
The histogram profiles, shown on the left, are generated from simulations of \num{e5} ions.
The ion range and strange at each implantation energy are shown on the right.
All measurements in this study correspond to average stopping depths \SI{\geq 30}{\nano\meter} below the crystal surface, well below where the topological surface state is expected to be important.
}
\end{figure}

As mentioned in \Cref{sec:experiment:measurement}, \ce{^{8}Li^{+}} implantation profiles in BTS were predicted using the \texttt{SRIM} Monte Carlo code~\cite{srim}.
At each implantation energy, stopping events were simulated for \num{e5} ions, with their resulting histogram representing the predicted implantation profile.
From the profiles shown in \Cref{fig:srim}, we calculate, in the nomenclature of ion-implantation literature, the range and straggle (\latin{i.e.}, the mean and standard deviation) at each simulated energy.
At the implantation energies used here (\SIrange{5}{28}{\kilo\electronvolt}), the incident \ce{^{8}Li^{+}} ions typically stop on average \SI{> 30}{\nano\meter} below the crystal surface, depths well below where the TSS is expected to be important.

\section{Helicity-resolved resonance spectra \label{sec:helicities}}

\begin{figure}
\centering
\includegraphics[width=1.0\columnwidth]{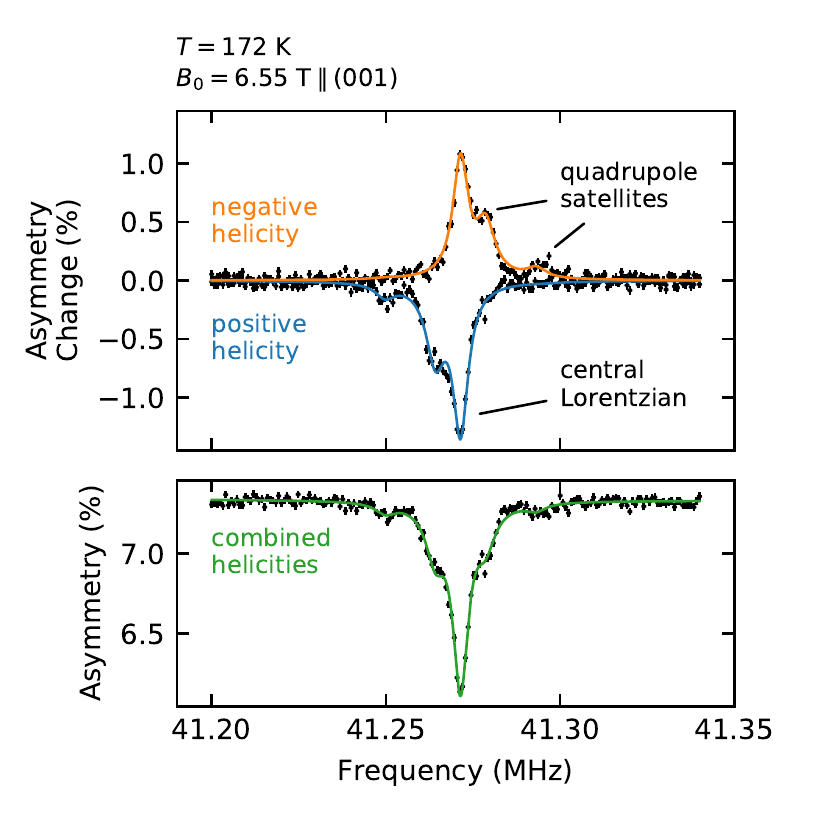}
\caption{ \label{fig:bts-1f-helicities}
Typical helicity-resolved \ce{^{8}Li} resonance spectra in \ce{Bi2Te2Se} with $\SI{6.55}{\tesla} \parallel (001)$, revealing the fine structure of the line.
Four quadrupole satellites, split asymmetrically in each helicity about a ``central'' Lorentzian line are evident.
Note that, in contrast to conventional NMR, the satellite amplitudes are determined primarily by the high degree of initial polarization~\cite{2014-MacFarlane-JPCS-551-012059}.
The solid coloured lines are global fits to a sum of five Lorentzians with positions given by \Cref{eq:nuq,eq:quadrupole,eq:angles} (see \Cref{sec:results:resonance} for further details).
Upon combining helicities, a nearly symmetric line about the resonance centre-of-mass, shown at the bottom inset, is obtained.
}
\end{figure}

Typical helicity-resolved resonance spectra are shown in \Cref{fig:bts-1f-helicities}, demonstrating the two key features in the line's fine structure.
A quadrupolar splitting on the order of several \si{\kilo\hertz}, clearly evidenced by the asymmetric shape about the resonance centre-of-mass in each helicity, can be associated with the outermost satellite lines.
Note that the satellite intensities are different from conventional NMR and are determined mainly by the high degree of initial polarization, which increases the relative amplitude of the outer satellites~\cite{2014-MacFarlane-JPCS-551-012059}, with their precise (time-average) values depending on the relaxation details~\cite{2009-Hossain-PB-404-914}.
Secondly, another significantly smaller quadrupolar frequency can be ascribed to a ``central'' Lorentzian-like line, analogous to what was observed in the structurally similar \ce{NbSe2}~\cite{2006-Wang-PB-374-239}. Note that there is no unshifted $m_{\pm 1/2} \leftrightarrow m_{\mp 1/2}$ magnetic sublevel transition, in contrast to spin $I=3/2$ \ce{^{7}Li}. The \rf\ amplitude dependence and the absence of other multiquantum lines indicate that it is also not a multiquantum transition.
Instead it must originate from the overlap of the four unresolved satellites with a small quadrupole splitting~\cite{2014-MacFarlane-PRB-90-214422}. Such a feature, in a noncubic layered crystal (see \Cref{fig:bts-unit-cell} in \Cref{sec:introduction}), is suggestive that this component originates from \ce{^{8}Li^{+}} within the vdW gap, where the magnitude of EFGs at interstices are minimized.
Resonances of the two helicities can be combined to give an overall average lineshape (see bottom inset in \Cref{fig:bts-1f-helicities}), whose evolution with temperature is shown in \Cref{fig:bts-1f-spectra} from \Cref{sec:results:resonance}.

\section{\ce{^{8}Li^{+}} sites \label{sec:discussion:site}}

Here we consider the stopping sites in more detail.
Generally, ion-implanted \ce{^{8}Li^{+}} occupies high-symmetry crystallographic sites that locally minimize its electrostatic potential.
This may include metastable sites that are not the energetic minimum, but have a significant potential barrier to the nearest stable site.
While these sites are characteristic of the isolated implanted ion, they may be related to the lattice location of \ce{Li^{+}} obtained by
chemical intercalation.

\begin{figure}
\centering
\includegraphics[width=1.0\columnwidth]{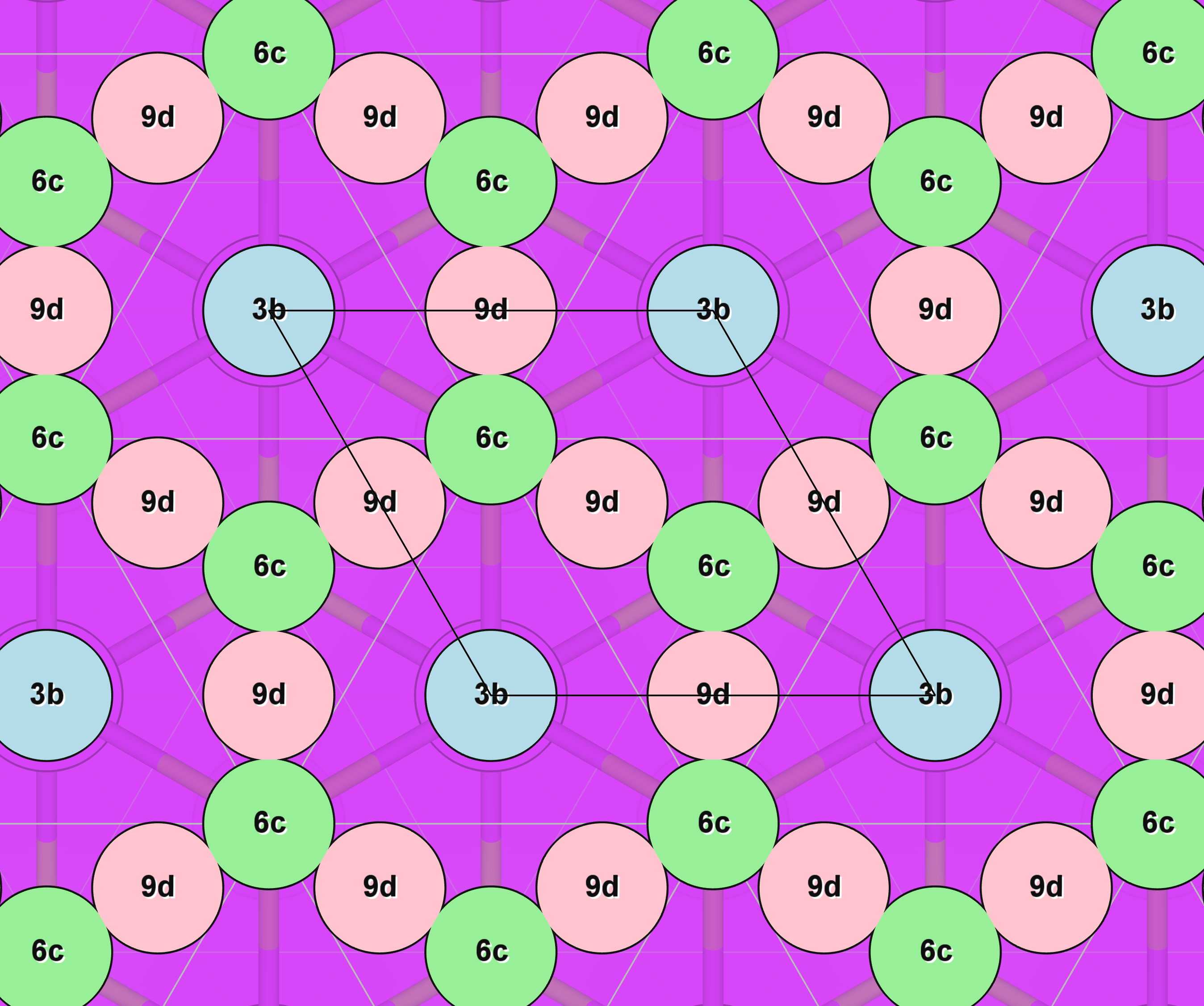}
\caption{ \label{fig:bts-vdwg-sites}
High-symmetry sites within the van der Waals gap of \ce{Bi2Te2Se}, shown as coloured circles, in a plane perpendicular to the trigonal $c$-axis.
The Wyckoff position is indicated for each site and the boundary of the unit cell is marked by solid black lines.
Neighbouring $3b$ (blue) sites, enclosed by \ce{Te} quasi-octahedra, are connected indirectly through $6c$ (green) sites with quasi-tetrahedral \ce{Te} coordination, and directly through $9d$ (pink) sites of lower symmetry.
The structure was drawn using \texttt{VESTA}~\cite{vesta}.
}
\end{figure}

BTS is structurally similar to the transitional metal dichalcogenides (TMDs) that consist of triatomic layers separated by a vdW gap between chalcogen planes.
This spacious interstitial region accommodates many types of extrinsic atoms and small molecules in the form of intercalation compounds~\cite{1976-Levy-CCCMLS}.
Similarly, a variety of dopants have been intercalated into the tetradymite \ce{Bi} chalcogenides, such as \ce{Cu}~\cite{1992-Maclachlan-JMS-27-4223}, \ce{Ag}~\cite{1968-Dibbs-JAP-39-2976}, \ce{Au}~\cite{2014-Shaughnessy-JAP-115-063705}, and \ce{Zn}~\cite{1960-Kuliev-SPSS-1-1076}.
Lithium has also been inserted into \ce{Bi} chalcogenides~\cite{1988-Paraskevopoulos-MSEB-1-147, 2009-Ding-JMC-19-2588, 2010-Bludska-JSSC-183-2813, 2011-Chen-JNR-13-6569}, but the precise sites for \ce{Li^{+}} in the vdW gap have not been determined~\cite{2010-Bludska-JSSC-183-2813}.
Note that in all these cases, intercalation is at the level of atomic \si{\percent}, so that intercalated species certainly interact (\latin{e.g.}, forming ``stage'' compounds~\cite{1976-Levy-CCCMLS}).

Based on this, we expect the lowest energy site for implanted \ce{^{8}Li^{+}} is within the vdW gap, similar to \ce{NbSe2}~\cite{2006-Wang-PB-374-239}.
Here, the EFGs are likely minimized, yielding a small quadrupole frequency, consistent with the unsplit component of the resonance in \Cref{fig:bts-1f-helicities}.
Within the vdW gap, several high-symmetry Wyckoff sites are available (see \Cref{fig:bts-vdwg-sites}): the quasi-octahedral $3b$ at $(0, 0, 1/2)$; the quasi-tetrahedral $6c$ at $(0, 0, 1/6)$; and the 2-fold coordinated $9d$ at $(1/2, 0, 1/2)$.
Here, the fractional coordinates correspond to the hexagonal unit cell in \Cref{fig:bts-unit-cell} from \Cref{sec:introduction}.
Neighbouring $3b$ sites are connected by direct paths through the $9d$ sites and indirect paths (\latin{i.e.}, dog-leg trajectories) passing through $6c$ sites.
The $3b$ site offers by far the largest coordination volume for interstitial \ce{^{8}Li^{+}} and it is reasonable that this is the preferred site in the vdW gap.
Indeed, preliminary density functional theory calculations confirm this assignment.

As indicated in \Cref{sec:results:resonance}, the low temperature resonances suggest a nearly 1:1 relative occupation of two sites with different EFGs.
Noting that, in contrast to the trilayers in TMDs, the QLs in BTS account a much larger volume fraction of the crystal, which leads us to consider possible interstitial sites therein.
While interstitial sites within the QL will be characterized by lower-symmetry and much larger EFGs, the most likely sites retain the trigonal rotation axis (\latin{e.g.}, $6c$ at $(0, 0, 1/3)$).
A simple point-charge model of isolated \ce{^{8}Li^{+}} in the BTS lattice, using ionic charges of \num[retain-explicit-plus]{+0.3} for \ce{Bi} and \num{-0.2} for \ce{Se}/\ce{Te}
($1/10$ their nominal values), gives $\nu_{q}$ for these sites that are within a factor \num{\sim 3} of the values for the $3b$ site in the vdW gap (\SI{\sim 2}{\kilo\hertz}), consistent with the difference required to explain the experimental spectra.
Note that, while the point charge model predicts nearly identical $|\nu_{q}|$ for all sites in the vdW gap, sites on the edges of the hexagons in \Cref{fig:bts-vdwg-sites} have the \emph{opposite} sign.

From the helicity-resolved resonances, the sign of the EFG does not change with temperature and, based on the $T$-independent $\nu_{q}$ and satellite amplitudes,
we suggest that this component corresponds to a fraction of implanted \ce{^{8}Li^{+}} that stops at a site within the QL, where it remains static over its lifetime.
Similarly, we ascribe the unsplit line to \ce{^{8}Li^{+}} stopped in the vdW gap, likely in the $3b$ site.
This component, in contrast, is dynamic above \SI{100}{\kelvin}, accounting for the growth in resonance amplitude and the $1/T_{1}$ maxima, which we consider in more detail in \Cref{sec:discussion:diffusion}.
Note that we find no evidence for a site change transition up to \SI{317}{\kelvin}, indicating a substantial energy barrier separates the two sites.

\end{document}